\documentclass[draft]{agujournal2019}
\usepackage{url} %this package should fix any errors with URLs in refs.
\usepackage{lineno}
\usepackage[inline]{trackchanges} %for better track changes. finalnew option will compile document with changes incorporated.
\usepackage{soul}
\usepackage{longtable}
\usepackage{multirow}
\usepackage{amsmath}
\UseRawInputEncoding
\nolinenumbers
\draftfalse

\journalname{JGR: Space Physics}

\begin{document}

%% ------------------------------------------------------------------------ %%
%  Title
%
% (A title should be specific, informative, and brief. Use
% abbreviations only if they are defined in the abstract. Titles that
% start with general keywords then specific terms are optimized in
% searches)
%
%% ------------------------------------------------------------------------ %%

% Example: \title{This is a test title}

\title{New Measurement of the Vertical Atmospheric Density Profile from Occultations of the Crab Nebula with X-Ray Astronomy Satellites Suzaku and Hitomi}

%% ------------------------------------------------------------------------ %%
%
%  AUTHORS AND AFFILIATIONS
%
%% ------------------------------------------------------------------------ %%

% Authors are individuals who have significantly contributed to the
% research and preparation of the article. Group authors are allowed, if
% each author in the group is separately identified in an appendix.)

% List authors by first name or initial followed by last name and
% separated by commas. Use \affil{} to number affiliations, and
% \thanks{} for author notes.
% Additional author notes should be indicated with \thanks{} (for
% example, for current addresses).

% Example: \authors{A. B. Author\affil{1}\thanks{Current address, Antartica}, B. C. Author\affil{2,3}, and D. E.
% Author\affil{3,4}\thanks{Also funded by Monsanto.}}

\authors{Satoru Katsuda\affil{1}, Hitoshi Fujiwara\affil{2}, Yoshitaka Ishisaki\affil{3,4}, Yoshitomo Maeda\affil{4}, Koji Mori\affil{5,4}, Yuko Motizuki\affil{6}, Kosuke Sato\affil{1}, Makoto S. Tashiro\affil{4,1}, and Yukikatsu Terada\affil{1,4}}

\affiliation{1}{Graduate School of Science and Engineering, Saitama University, 255 Shimo-Ohkubo, Sakura, Saitama 338-8570, Japan}
\affiliation{2}{Faculty of Science and Technology, Seikei University, Tokyo, Japan}
\affiliation{3}{Department of Physics, Tokyo Metropolitan University, 1-1 Minami-Osawa, Hachioji, Tokyo 192-0397, Japan}
\affiliation{4}{Institute of Space and Astronautical Science (ISAS), Japan Aerospace Exploration Agency (JAXA), 3-1-1 Yoshinodai, Chuo-ku, Sagamihara, Kanagawa, 252-5210, Japan}
\affiliation{5}{Department of Applied Physics and Electronic Engineering, University of Miyazaki, 1-1, Gakuen Kibanadai-nishi, Miyazaki 889-2192, Japan}
\affiliation{6}{RIKEN Nishina Center, Hirosawa 2-1, Wako 351-0198, Japan}

%\affiliation{=number=}{=Affiliation Address=}
%(repeat as many times as is necessary)

%% Corresponding Author:
% Corresponding author mailing address and e-mail address: katsuda@phy.saitama-u.ac.jp

% (include name and email addresses of the corresponding author.  More
% than one corresponding author is allowed in this LaTeX file and for
% publication; but only one corresponding author is allowed in our
% editorial system.)

% Example: \correspondingauthor{First and Last Name}{email@address.edu}

\correspondingauthor{Satoru Katsuda}{katsuda@phy.saitama-u.ac.jp}

%% Keypoints, final entry on title page.

%  List up to three key points (at least one is required)
%  Key Points summarize the main points and conclusions of the article
%  Each must be 140 characters or fewer with no special characters or punctuation and must be complete sentences

% Example:
% \begin{keypoints}
% \item	List up to three key points (at least one is required)
% \item	Key Points summarize the main points and conclusions of the article
% \item	Each must be 140 characters or fewer with no special characters or punctuation and must be complete sentences
% \end{keypoints}

\begin{keypoints}
\item Combined O and N densities at altitudes 70--200\,km are measured from Earth occultations of the Crab Nebula using X-ray astronomy satellites, Suzaku and Hitomi.  
\item The vertical density profile is in general agreement with a predicted profile from the NRLMSISE-00 model, except for altitudes 70--110\,km in which the density is significantly smaller than the prediction by the NRL model. 
\item This density deficit could be due to either long-term radiative cooling of the upper atmosphere or imperfect modeling.  
\end{keypoints}

%% ------------------------------------------------------------------------ %%
%
%  ABSTRACT and PLAIN LANGUAGE SUMMARY
%
% A good Abstract will begin with a short description of the problem
% being addressed, briefly describe the new data or analyses, then
% briefly states the main conclusion(s) and how they are supported and
% uncertainties.

% The Plain Language Summary should be written for a broad audience,
% including journalists and the science-interested public, that will not have 
% a background in your field.
%
% A Plain Language Summary is required in GRL, JGR: Planets, JGR: Biogeosciences,
% JGR: Oceans, G-Cubed, Reviews of Geophysics, and JAMES.
% see http://sharingscience.agu.org/creating-plain-language-summary/)
%
%% ------------------------------------------------------------------------ %%

%% \begin{abstract} starts the second page

\begin{abstract}
We present new measurements of the vertical density profile of the Earth's atmosphere at altitudes between 70 and 200\,km, based on Earth occultations of the Crab Nebula observed with the X-ray Imaging Spectrometer onboard Suzaku and the Hard X-ray Imager onboard Hitomi.  X-ray spectral variation due to the atmospheric absorption is used to derive tangential column densities of the absorbing species, i.e., N and O including atoms and molecules, along the line of sight.  The tangential column densities are then inverted to obtain the atmospheric number density.  The data from 219 occultation scans at low latitudes in both hemispheres from September 15, 2005 to March 26, 2016 are analyzed to generate a single, highly-averaged (in both space and time) vertical density profile.  The density profile is in good agreement with the NRLMSISE-00 model, except for the altitude range of 70--110\,km, where the measured density is $\sim$50\% smaller than the model.  Such a deviation is consistent with the recent measurement with the SABER aboard the TIMED satellite \cite{2020Atmos..11..341C}.  Given that the NRLMSISE-00 model was constructed some time ago, the density decline could be due to the radiative cooling/contracting of the upper atmosphere as a result of greenhouse warming in the troposphere.  However, we cannot rule out a possibility that the NRL model is simply imperfect in this region.  We also present future prospects for the upcoming Japan-US X-ray astronomy satellite, XRISM, which will allow us to measure atmospheric composition with unprecedented spectral resolution of $\Delta E \sim 5$\,eV in 0.3--12\,keV.

\end{abstract}

%\section*{Plain Language Summary}
%[ enter your Plain Language Summary here or delete this section]

%% ------------------------------------------------------------------------ %%
%
%  TEXT
%
%% ------------------------------------------------------------------------ %%

%%% Suggested section heads:
\section{Introduction}

The neutral density in the lower thermosphere, defined as an altitude range of 70--200\,km in this paper, is essential to estimate the global meteoric mass input and to derive the momentum flux from tropospheric gravity waves.  Also, the lower thermosphere is thought to be sensitive to the climate change due to greenhouse gases; global cooling would occur in the upper atmosphere in conjunction with global warming in the troposphere due to long-term increase of greenhouse gas concentrations, which was first pointed out by \citeA{1989GeoRL..16.1441R}.  In contrast to the troposphere, where CO$_2$ is optically thick and traps IR radiation emitted by the Earth's surface, in the stratosphere and above, CO$_2$ is optically thin and emits infrared radiation to space, causing cooling in the middle and upper atmosphere.  At a given pressure, this cooling must result in a density increase.  However, in the combination of the thermal shrinking, atmospheric density at a given height decreases in the middle and upper atmosphere.

Measurements of neutral densities in the lower thermosphere are still scarce.  Neutral densities are largely measured by means of in-situ instruments onboard sounding rockets (80--100\,km) and satellites ($\sim$400\,km), and thus the 100--300\,km altitude is left as the ``thermospheric gap" \cite{2011JGRA..11611306O}.  Although there are some early density measurements based on occultations of the Sun in ultraviolet wavelengths \cite<e.g.,>{1968JGR....73.5798N,1972Sci...176..793H,1974JGR....79.4757A,1992P&SS...40.1153M,1993JGR....9817607A,2007JGRD..11216308L}, these studies mainly focused on molecular oxygen which is a minor species in the lower thermosphere, leaving large uncertainties on the total mass density.  So far, there are very limited remote-sensing measurements to fill this gap \cite<e.g., >{2015E&SS....2....1M}.  

Atmospheric occultations of X-ray astronomical sources offer a unique opportunity to measure the neutral density in the lower thermosphere.  This technique was first demonstrated/published by \citeA{2007JGRA..112.6323D} who analyzed data during atmospheric occultations of the Crab Nebula and Cygnus X-2 using ARGOS/USA and RXTE/PCA.  We note that there are a few interesting X-ray measurements of atmospheric structures at solar planets and their satellites.  The atmospheric thickness of Titan (i.e., the Saturn's largest satellite) was measured from the transit of the Crab Nebula on 2003-01-05 \cite{2004ApJ...607.1065M}.  Very recently, an atmospheric density profile at Mars was measured by occultations of $\sim$10\,keV X-rays from Scorpius X-1, using the SEP instrument on the MAVEN spacecraft \cite{2020GeoRL..4788927R}.  The advantage of the X-ray occultation method is that it is independent of the chemical, ionization, and excitation processes, because X-rays are directly absorbed by inner K-shell and L-shell electrons, so that X-rays see only atoms (within molecules).  In other words, the X-ray occultation method cannot distinguish between atoms and molecules, but allows us to measure the total neutral density without the complexity involved with modeling absorption processes.   

By analyzing occultations of the Crab Nebula taken in November 2005, \citeA{2007JGRA..112.6323D} reported that 50\% and 15\% smaller densities at altitudes 100--120\,km and 70--90\,km, respectively, than the empirical density model, i.e., the Naval-Research-Laboratory's-Mass-Spectrometer-Incoherent-Scatter-Radar-Extended model \cite<NRLMSISE-00:>{2002JGRA..107.1468P}.  This deviation is consistent with the density decrease due to greenhouse cooling in the upper atmosphere, given that a few decades have past after the NRL model was constructed.  There are a number of theoretical studies of the upper-atmosphere cooling induced by increasing greenhouse gases.  \citeA{1998AnGeo..16.1501A} gave a nice review on this topic.  Although they argued that there still remain uncertainties in model estimates especially in the mesosphere and lower thermosphere region, multiple studies suggested that doubling CO$_2$ experiment causes a density decrease of 50\% at $\sim$100\,km \cite{1989GeoRL..16.1441R,1992P&SS...40.1011R,1998AnGeo..16.1501A}.  At a glance, this appears to be consistent with the observation by \citeA{2007JGRA..112.6323D}.  It should be noted, however, that the increasing rate of CO$_2$ in the atmosphere was measured to be $\sim$12\% during 1960--1992 \cite<e.g.,>{1996JATP...58.1673K}, which should result in a much lower rate in the density decrease than that observed.  Therefore, the remarkable density difference between the data \cite{2007JGRA..112.6323D} and the NRL model is not yet convincing.  In this context, it is important to cross-check the density structure from other observations, to determine systematic errors in the retrieval, and to distinguish between greenhouse gases and solar effects by monitoring over times longer than the solar cycle.  Therefore, more observations are required to firmly conclude (or exclude) this interpretation.  

Here, we present a new density measurement in the lower thermosphere region, based on occultations of the Crab Nebula observed with the Japanese X-ray astronomy satellites Suzaku \cite{2007PASJ...59S...1M} and Hitomi \cite{2018JATIS...4b1402T}.  The combination of the X-ray Imaging Spectrometer (XIS, sensitive in 0.2--12\,keV: \citeA{2007PASJ...59S..23K}) and the Hard X-ray Imager (HXI, sensitive in 5--80\,keV: \citeA{2018JATIS...4b1410N}) aboard Suzaku and Hitomi, respectively, allows us to investigate the density structure in a range of altitude between 70\,km and 200\,km.  The paper is organized as follows.  Section~2 describes occultation data of the Crab Nebula acquired with Suzaku and Hitomi. Section~3 describes analysis of the data, and shows the retrieved vertical density distribution.  Section~4 discusses the results, and provides future prospects.  Section~5 gives conclusions of this paper.

\section{Observations}

We analyze data during Earth occultations of the Crab Nebula, obtained with the X-ray astronomy satellites Suzaku and Hitomi.  The Crab Nebula and the Crab pulsar were created in a supernova explosion observed in 1054~AD.  Various X-ray satellites have used the Crab Nebula as a standard candle to perform their calibrations in the past, because it is one of the brightest sources in the X-ray sky, and its intensity is nearly constant and the spectral shape is a simple, featureless power-law.  

Both Suzaku and Hitomi carried multiple instruments with different capabilities.  Of these, we focus on Suzaku/XIS and Hitomi/HXI, because they are complementary in terms of the energy coverage: 0.2--12\,keV for the XIS and 5--80\,keV for the HXI, and also they have time and spatial resolutions sufficient for our purpose.

\begin{figure}
\includegraphics[width=35pc,angle=0]{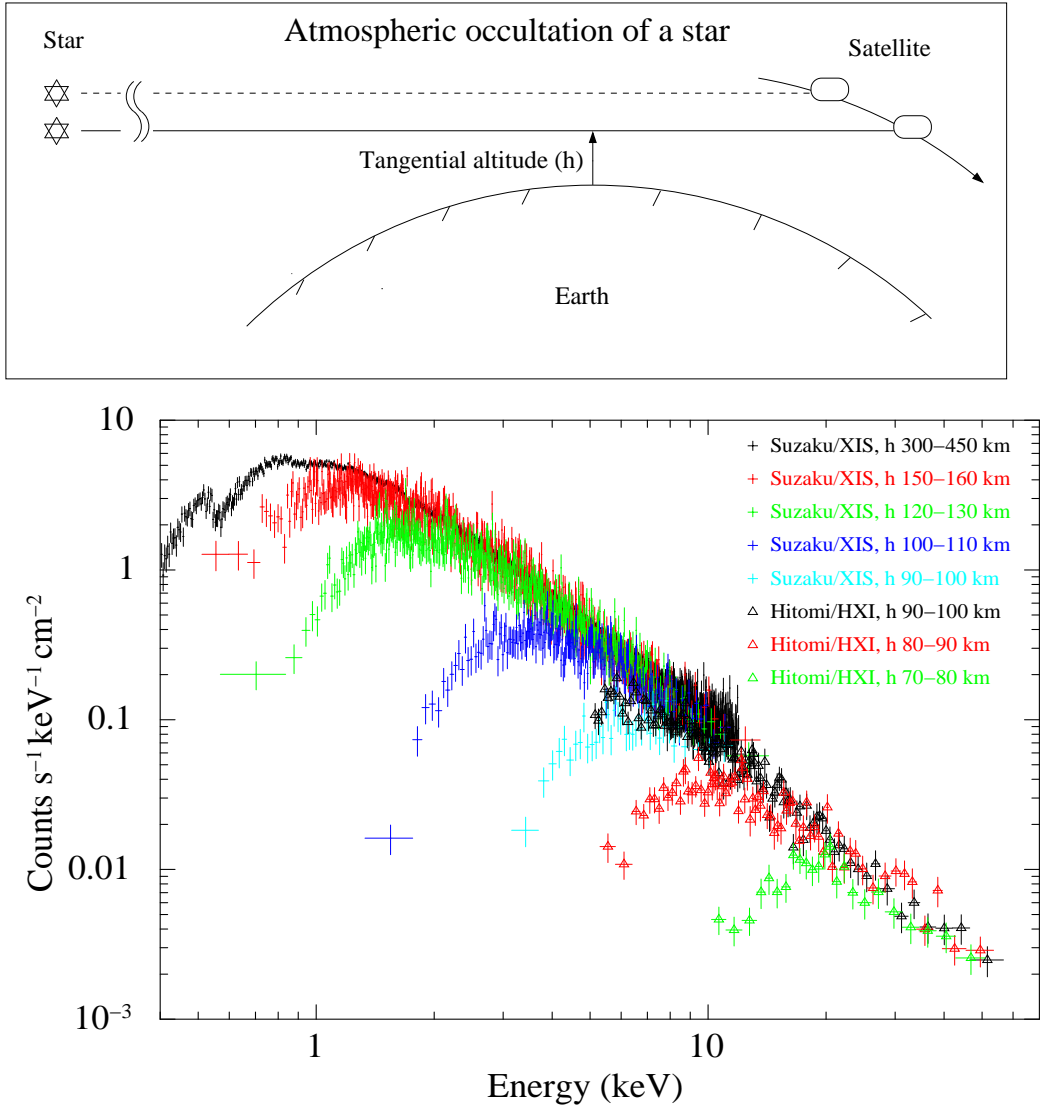}
\caption{Upper panel: Geometry of Earth's atmospheric occultation of a star.  As the satellite proceeds, the source sets behind the Earth's atmosphere.  Lower panel: X-ray spectral variation of the Crab Nebula during the occultation.  The data are taken with the XIS and HXI covering 0.4--12\,keV and 4--70\,keV, respectively.  Spectra free from the atmospheric absorption are shown in black.  As the occultation progresses, the X-ray photons are gradually absorbed from the low-energy side due to the increasing atmospheric density with the decreasing tangential altitude.}
\label{fig:spec_variation}
\end{figure}

As illustrated in the upper panel of Figure~\ref{fig:spec_variation}, X-ray sources appear to set and rise behind the Earth every orbit.  This is because the two X-ray astronomy satellites are in low Earth orbits, i.e., $\sim$560\,km (Suzaku) and $\sim$575\,km (Hitomi), and also the attitude of the satellites is fixed during each observation.  The lower panel of Figure~\ref{fig:spec_variation} demonstrates how the X-ray spectrum changes during an atmospheric occultation of the Crab Nebula, where the spectra are the sums of all the setting data listed in Table~\ref{tab:data}.  The data in tangential altitudes 300--450\,km are free from the atmospheric attenuation, where the tangential altitude ($h$) is defined as the altitude of closest approach to the Earth surface along the telescope line of sights (see also Figure~\ref{fig:spec_variation}).  As the occultation progresses (decreasing tangential altitude), the X-ray intensity gradually decreases from the low-energy band.

In principle, all X-ray sources can be our targets, but in practice, the targets are very limited for the following reasons.  First, the targets must be very bright, given that typical occultations take only $\sim$30\,s, which needs to be further divided into shorter time bins to obtain a vertical density distribution.  Second, the target's spectrum and intensity should be constant during the occultation to avoid confusion between intrinsic source variations and atmospheric absorption.  Third, the targets are preferably point sources for simplicity of the analysis.  

The source brightness also has an important impact on the time resolution of Suzaku/XIS.  The large angular velocity of the telescope, 0.06$^\circ$\,s$^{-1}$, for both satellites, as well as the short transition period from unattenuation to full attenuation, $\sim$30\,s, require time resolution of the order of second or less to resolve vertical atmospheric density structures.  The time resolution of the XIS in a normal mode is 8\,s.  This corresponds to $\sim$0.5$^\circ$ ($\sim$0.06$^\circ$\,s$^{-1}$$\times$8\,s) or $\sim$20\,km at the altitude of 100\,km, and is too coarse for our diagnostic purposes.  Therefore, we decided to focus on data taken in a special mode, i.e., 0.1-s burst mode, in which the exposure time is limited to 0.1\,s for each 8-s sampling.  This mode was prepared to avoid pile-up effects occurring for super-bright sources \cite<>[for more details]{2012PASJ...64...53Y}.  As for Hitomi/HXI, the time resolution does not matter, because the time-tagging system for the HXI is event-by-event with excellent resolution of 25.6\,$\mu$s.

The Crab Nebula is an unrivaled source to meet all the criteria above.  Also, it was usually observed with the XIS in the 0.1-s burst mode, and thus is an ideal source for the measurement of the atmospheric density profile.  In addition, there are numerous Suzaku data for the Crab Nebula, because it was one of the important calibration sources for Suzaku, as is usual for X-ray astronomy satellites.  This is another advantage with the Crab Nebula.  Although the Crab Nebula is not a point source, its angular size of $R\sim$1$^\prime$ is smaller than the half-power diameter (HPD) of the X-ray telescopes on Suzaku \cite<HPD$\sim$2$^{\prime}$:>{2007PASJ...59S...9S} and Hitomi \cite<HPD$\sim$1.7$^{\prime}$:>{2018JATIS...4a1212M}.  Therefore, it can be considered nearly a point source for both satellites.   Note that the angular extent of 1$^\prime$ from the satellite's view point corresponds to a physical extent of 0.7\,km at the Earth's limb.  This can be regarded as the minimal altitude resolution of our study.

Table~\ref{tab:data} summarizes basic information about all occultations of the Crab Nebula analyzed in this paper.  The second column gives the time at which the tangential altitude becomes 100\,km.  We computed the tangential altitude, by making use of the python package {\tt pyephem} (available at {\tt https://pypi.org/project/pyephem/}).  In this procedure, we need to know satellites' positions.  Although archival data for both Suzaku and Hitomi contain information about their locations, we recalculated satellites' positions by ourselves to improve the accuracy, using the {\tt pyephem} package into which we gave orbital elements taken from NORAD two-line elements (TLEs: Data for both Suzaku and Hitomi are accessible by sending special requests at {\tt http://celestrak.com/NORAD/archives/request.php}).  The third column indicates the Earth's latitude and longitude at the tangent point.  The forth column lists the local time at the tangent point.  We note that most observations are taken in the latitude from $-$20$^\circ$ to $+$40$^\circ$.  This is because the inclination angles of both satellites are 31$^{\circ}$, and the telescope is 22$^\circ$ inclined to the north, toward the Crab Nebula.  

Suzaku observed the Crab Nebula almost every spring and autumn for ten years.  Many of these observations captured Crab at off-axis positions for calibration purposes.  For simplicity of our analysis, we selected observations aiming at the Crab Nebula at on-axis positions.  However, we allow tolerance of $\sim$3$^\prime$ offset in Decl.\ (north-south) direction.  This is because the Crab Nebula sets and rises roughly from the east-west (i.e., R.A.) direction, and thus the north-south offset does not affect setting and rising times by much.  We also discarded short-exposure (duration less than 10\,ks) observations for Suzaku to improve the efficiency of our analysis.

\section{Analysis and Results}

\begin{figure}
\includegraphics[width=35pc,angle=0]{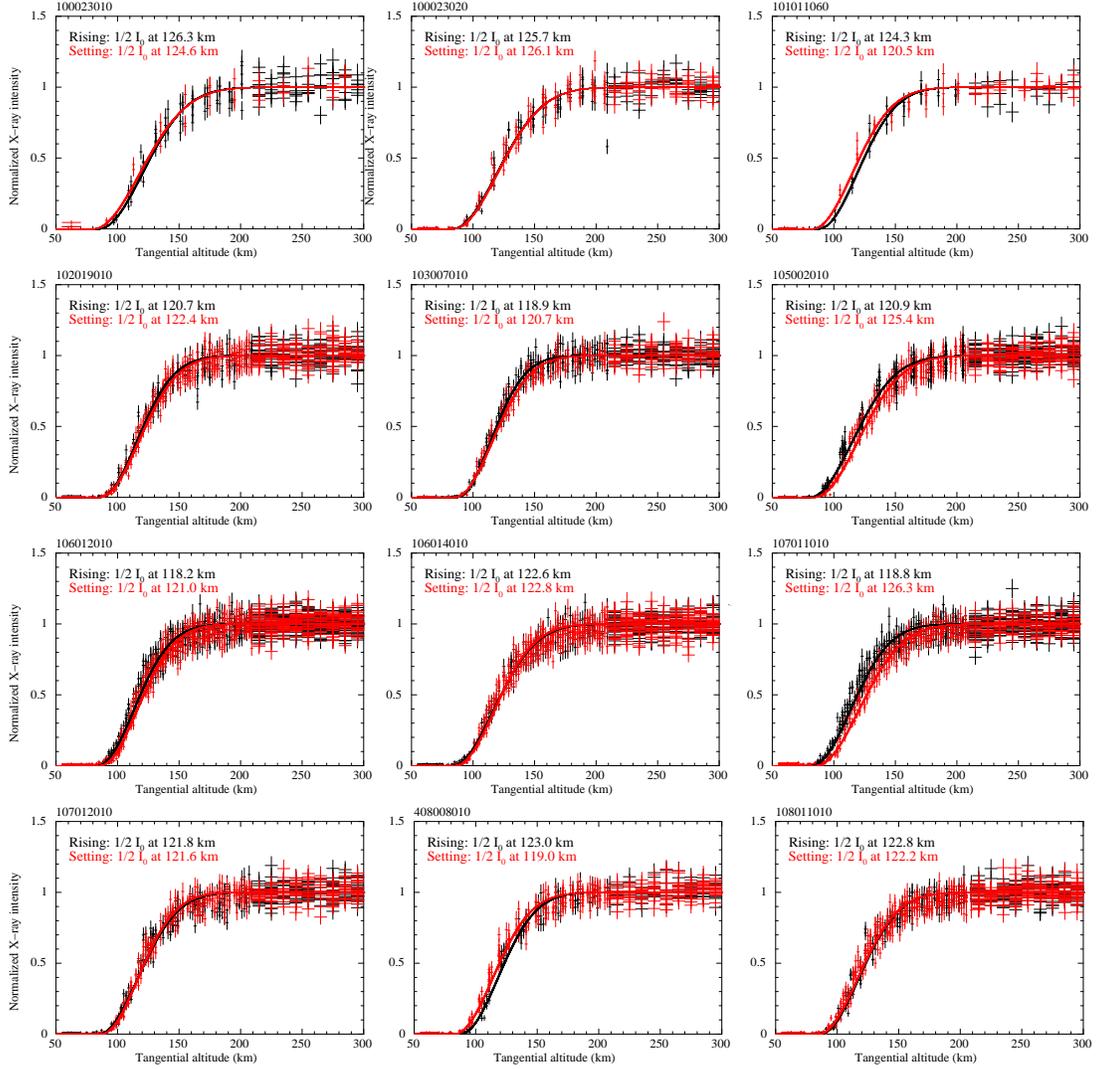}
\caption{Occultation light curves in 0.5--10\,keV for Suzaku/XIS0, XIS1, and XIS3.  Black and red are responsible for rising and setting, respectively.  The intensities are normalized at the unattenuated level.  These profiles are fitted with a phenomenological model shown in blue, from which we computed an altitude at which the flux becomes half of the unattenuated level, which is indicated in the upper left corner of each panel.}
\label{fig:suzaku_lc}
\end{figure}

\begin{figure}
\includegraphics[width=35pc,angle=0]{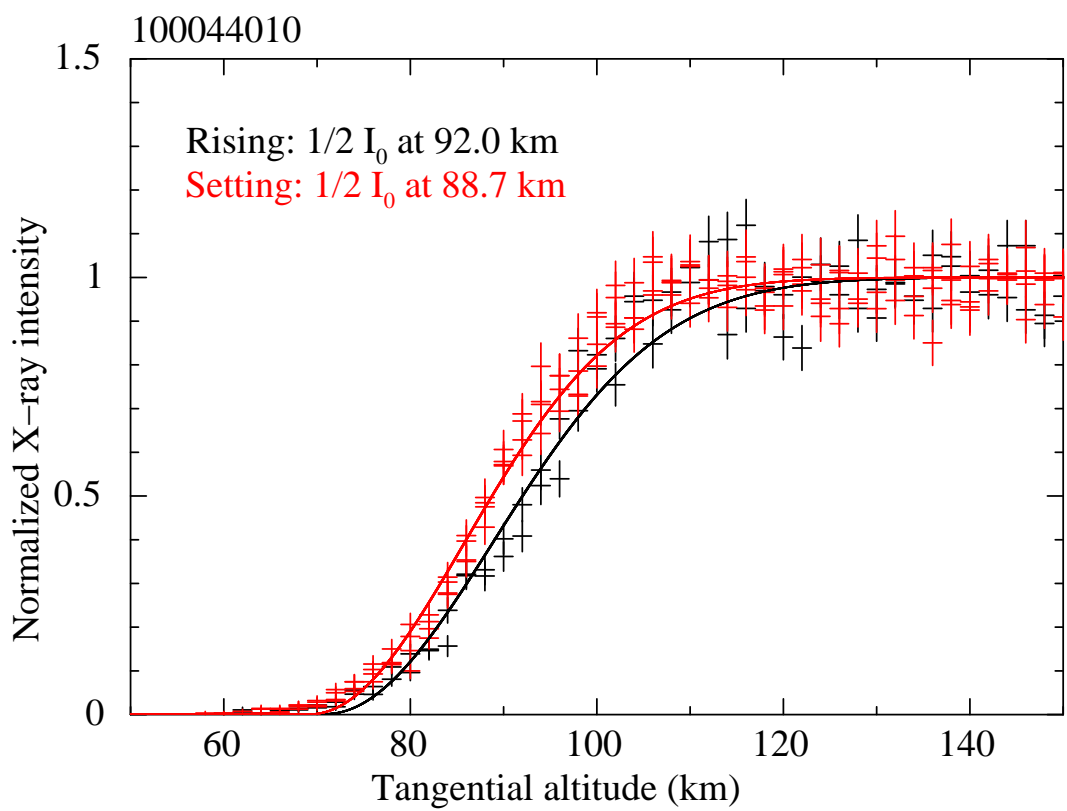}
\caption{Same as Figure~\ref{fig:hitomi_lc} but for all the data taken with Hitomi/HXI1 and HXI2.  The energy range used is 5--50\,keV.}
\label{fig:hitomi_lc}
\end{figure}

When X-ray sources set behind (or rise from) the Earth's atmosphere, we see rounded, not sharp-edged, intensity profiles.  Figures~\ref{fig:suzaku_lc} and \ref{fig:hitomi_lc} exhibit all the occultation light curves obtained with Suzaku/XIS and Hitomi/HXI, respectively, whose detailed information is listed in Table~\ref{tab:data}.  In each panel, there are multiple data from different instruments, i.e., XIS0, XIS1, and XIS3 for Suzaku and HXI1 and HXI2 for Hitomi.  The intensity is normalized at the unattenuated level ($I_0$) for each observation epoch.  The gradual increase/decrease of the X-ray intensity with the tangential altitude clearly shows the effect of atmospheric absorption.

We fit these profiles with a phenomenological model consisting of Gaussian plus constant components as below:\\
\begin{equation}
I = \begin{cases}
I_0 - I_0 \exp(-\frac{(x - x_c)^2}{2\sigma^2}) + b & (x > x_c)\\
b~~~~({\rm otherwise}),\\
\end{cases}
\end{equation}
where $I$ and $x$ are the X-ray intensity and the altitude, respectively.  Taking $I_0$, $x_c$, $\sigma$, and $b$ as free parameters, we fit the individual profiles.  The best-fit curves are shown in black (rise) and red (set) in Figures~\ref{fig:suzaku_lc} and \ref{fig:hitomi_lc}.  Based on the best-fit model, we calculated an altitude at which $I = 1/2 \times I_0$.  The half-flux altitudes are given in the upper-left corner in each panel.  The fact that Hitomi/HXI's value is significantly smaller than those for Suzaku/XIS is because Hitomi/HXI is more sensitive to the higher energy band than Suzaku/XIS.  

The two (set and rise) groups in each observation epoch are shifted with respect to each other along the x-axis by 0.2--7.5\,km.  This variation seems random; no seasonal trend, long-term trend, nor latitude dependence can be found.  This implies that the shift does not reflect real density differences, but is caused by instrumental/artificial effects such as uncertainties on the satellite's position or time assignments to the data.  As for the satellite's position accuracy, orbit predictions from NORAD TLEs have been shown to be accurate to within $\pm$0.5\,km \cite{2018Ap&SS.363...31X}.  Given the satellite's orbital velocity of 7.5\,km\,s$^{-1}$ expected for both Suzaku and Hitomi, a misplacement of the satellite by 1\,km could cause $\pm$0.13\,s shifts between the setting and rising timings.  This corresponds to $\sim$0.5\,km shifts for the tangential altitude, so that the satellite's position error seems to be insufficient to fully explain the discrepancy between the set and rise profiles.   As for the time assignments, the absolute timing accuracies for Hitomi/HXI and Suzaku/XIS (in the normal mode with a 0.1-s burst option) were verified in orbit at $\sim$300\,$\mu$s \cite{2018JATIS...4b1402T} and 24\,ms \cite{2010ecsa.conf..412M,2008PASJ...60S..25T}, respectively.  These are much less than the uncertainty due to the satellites' locations.  Therefore, the instrumental effects examined above seem not to be major sources to shift the set and rise profiles.  Because we do not have any more good ideas for potential systematic uncertainties, we attribute the shift to the difference of the atmospheric densities, and we measured density profiles for setting and rising, separately, in the following analysis.

\begin{figure}
\includegraphics[width=35pc,angle=0]{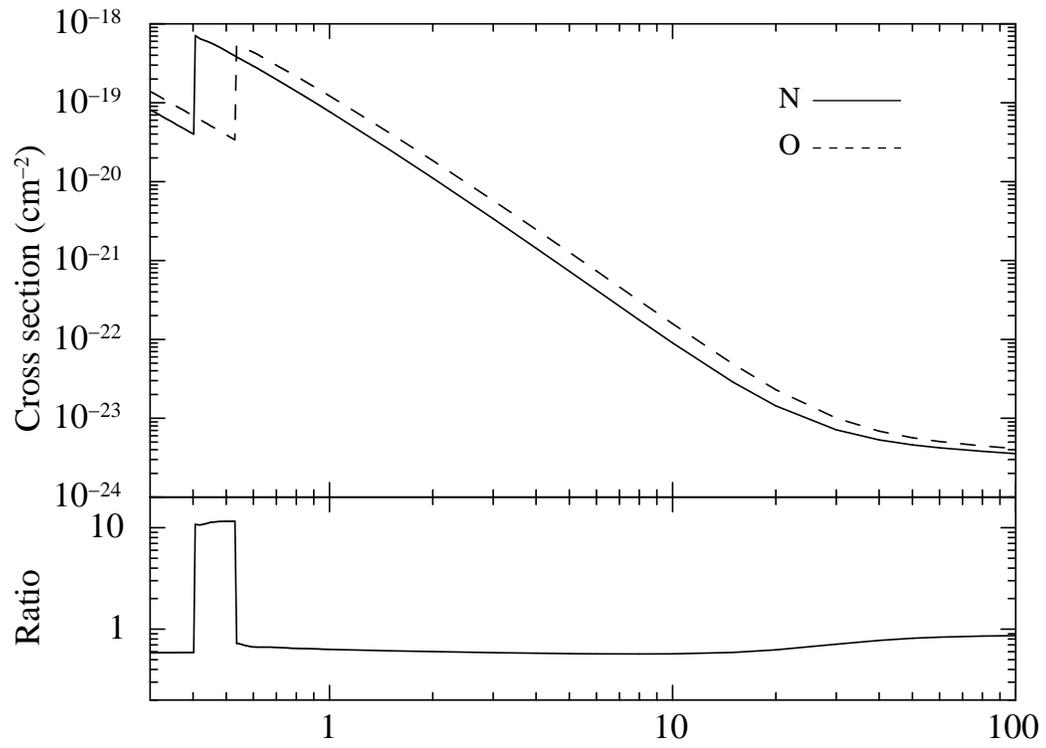}
\caption{Upper panel: X-ray cross sections for N and O. Lower panel: Cross-section ratio of N to O.  The ratio is nearly constant except for the K-shell edges at $\sim$0.5\,keV.  }
\label{fig:Xsect}
\end{figure}

A standard method to retrieve density profiles from the occultation technique has been based on the Beer's law:
\begin{equation}
I(\lambda) = I_0(\lambda) \exp(-\tau),
\end{equation}
where $I_0(\lambda)$ is the unattenuated source intensity, and $\tau$ is the optical depth at a specific altitude ($h$).  In the upper atmosphere in altitudes above 70\,km, it is reasonable to assume that $\tau = N_{\rm N} \times \sigma_{\rm N} + N_{\rm O} \times \sigma_{\rm O}$, where $N_{\rm N}$ and $N_{\rm O}$ are the N and O column densities along the line of sight, $\sigma_{\rm N}$ and $\sigma_{\rm O}$ are photo-absorption cross sections for the two elements.  Because cross sections are known, it is possible to derive N and O column densities.  If we have data at several wavelengths, it would be possible to measure N and O densities, separately.  It should be noted, however, that as shown in Figure~\ref{fig:Xsect} the energy dependence of cross sections of N and O are similar with each other, with a clear difference only around 0.5\,keV, where K-shell edges of these elements are present (0.4\,keV and 0.52\,keV for N and O).  Unfortunately, the data from both Suzaku/XIS and Hitomi/HXI are not very sensitive to this energy band.  The XIS is capable of detecting photons at this energy band, but its energy resolution of 50--60\,keV (FWHM) below 1\,keV (cf., {\tt https://heasarc.gsfc.nasa.gov/docs/astroe/prop\_tools/suzaku\_td/node10.html}) is just half of the energy difference between N and O K edges.  With this moderate energy resolution, a large amount of photons are required to resolve N and O K-edges.  The Crab Nebula's X-ray emission around 0.5\,keV is heavily ($\sim$90\%) absorbed by the interstellar medium, so that relatively weak emission can reach the Earth.  Therefore, it is difficult to measure N and O column densities separately from our data sets.  

It is possible to measure the combined N$+$O column densities, if we assume their relative abundances.  However, there still remains a difficulty in deriving a density profile from the Beer's-law method.  Given that the X-ray spectrum from the Crab Nebula is a pure power-law continuum, we need to consider the energy dependence of cross sections, even if we partition the spectrum into several energy bins.  This makes analysis procedures more complicated than cases in which emission/absorption lines are analyzed.

In contrast, it is technically easy to take account of the energy dependence of the photo-absorption cross sections, if we utilize absorption models equipped in the XSPEC package \cite{1996ASPC..101...17A}, which is a standard spectral analysis software in the X-ray astronomy.  We thus decided to adopt the spectral-fitting method to derive atmospheric densities.  Due to the limited photon statistics and sampling rates in one occultation scan especially for Suzaku, we accumulated all occultations in each Obs.ID, after filtering by rising and setting.  We further combine short-exposure observations, i.e., Obs.IDs~100023010, 100023020, and 101011060, to improve the effective sampling rate.  To increase the photon statistics, we combine data from Suzaku's XIS0, XIS1, and XIS3, and Hitomi's HXI1 and HXI2.  As a result, we obtained 10 observation epochs.  For each epoch, we extracted X-ray spectra by slicing altitudes of 90--210\,km (Suzaku) and 70--100\,km (Hitomi) into 10-km bins.  Note that we oversampled the altitudes 105--145\,km, where Suzaku/XIS is the most sensitive to the atmospheric absorption and the exponent of the atmospheric density structure changes.

For Suzaku/XIS, the brightest region of the Crab Nebula, namely the core region around the Crab pulsar, is subject to the so-called pileup effects.  For very bright sources, more than one photons strike the same CCD pixel or one of its immediate neighbors during the exposure.  When two photons with energies of $E_1$ and $E_2$ fall onto the same pixel, it is impossible to distinguish them from one with an energy of $E_1 + E_2$.  This causes a lower flux and a harder spectral shape than real.  Therefore, we excluded a central circular region (a radius of 40$^{\prime\prime}$) that include the Crab pulsar, where a pileup fraction was reported to exceed 3\% \cite{2012PASJ...64...53Y}.
% 0.1 deg --> $\sim$2\,km

\begin{figure}
\includegraphics[width=35pc,angle=0]{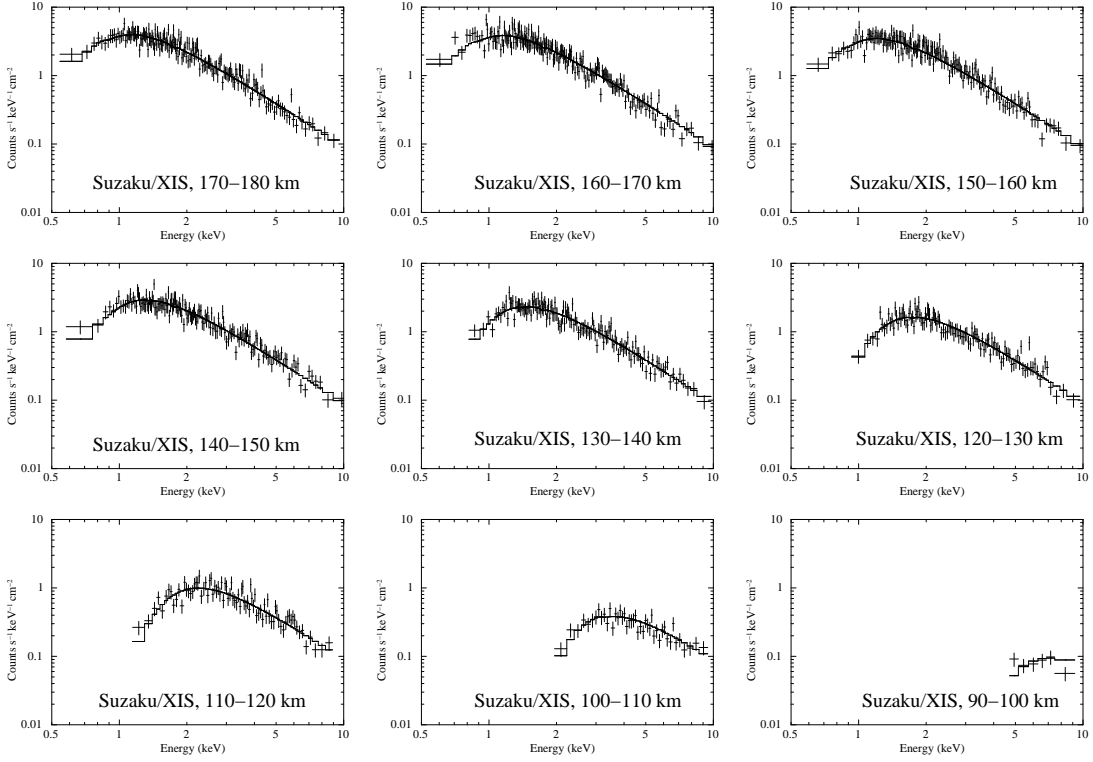}
\caption{Example X-ray spectra obtained with the XIS (Obs.ID~106014010; all setting data in this observation are combined together).  The data are fitted with the emission model from the Crab Nebula, which is multiplied by the Earth's atmospheric absorption.}
\label{fig:suzaku_spec}
\end{figure}

\begin{figure}
\includegraphics[width=35pc,angle=0]{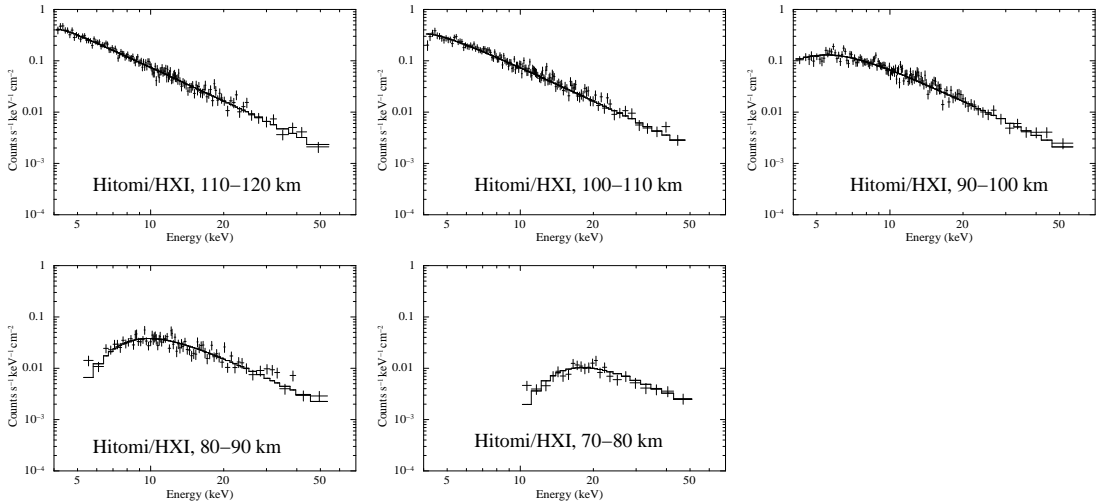}
\caption{Same as Figure~\ref{fig:suzaku_spec} but for Hitomi.}
\label{fig:hitomi_spec}
\end{figure}

Figures~\ref{fig:suzaku_spec} and \ref{fig:hitomi_spec} exhibit example X-ray spectra from Suzaku/XIS (Obs.ID~106014010) and Hitomi/HXI, respectively.  Clearly, the X-ray intensity decreases from the low-energy band as the line-of-sight goes deeper into the atmosphere.  We first determined an unattenuated spectrum, i.e., the source spectrum at the top of Earth's atmosphere, for each observation epoch.  To this end, we fitted X-ray spectra taken during high tangential altitudes of 300--450\,km with a conventional emission model for the Crab Nebula.  The model consists of {\tt TBabs} $\times$ power-law, where {\tt TBabs} is the interstellar absorption model \cite{2000ApJ...542..914W} in XSPEC \cite{1996ASPC..101...17A}.  We checked that the best-fit parameters in all epochs are in reasonable agreement with previous measurements \cite<e.g.,>{2018JATIS...4b1409H}.  

Then, we fitted lower-altitude spectra with the unattenuated spectral model which is taken account of the atmospheric absorption.  For the atmospheric absorption, we adopted the {\tt vphabs} with the photoionization cross-sections by \citeA{1996ApJ...465..487V}.  As described in \citeA{2007JGRA..112.6323D}, the accuracy of absolute cross sections is known to be better than 5\%, which is smaller than a typical statistical uncertainty on each measurement (cf., Tables~\ref{tab:NOcol_suzaku} and \ref{tab:NOcol_hitomi}).  In this fitting procedure, all parameters related to the Crab Nebula, i.e., the hydrogen column density for the interstellar absorption, the photon index and normalization of the power-law component, are fixed at the best-fit values obtained for each Obs.ID.  In the atmospheric absorption component, we consider N, O, and Ar, with other elements fixed to zero.  It is difficult to separately measure N and O column densities even with the spectral-fitting method due to the almost indistinguishable energy dependences of the N and O cross sections in the energy range of our interest.  Therefore, we fixed relative abundances of O/N and Ar/N to those expected in the NRLMSISE-00 model at each altitude of interest \cite{2002JGRA..107.1468P}.  Ar is a relatively minor element compared with N and O; the density of Ar is expected to be much smaller (less than 1\%) than the N+O density in the lower thermosphere according to the NRLMSISE-00 model.  Thus, Ar is not included in the following discussion, although we considered a little contribution from Ar in our spectral analyses.

In the spectral fitting, the setting and rising data were analyzed individually, given that their half-flux altitudes are often different from one another as shown in Figures~\ref{fig:suzaku_lc} and \ref{fig:hitomi_lc}.  Example fitting results are shown in Figures~\ref{fig:suzaku_spec} and \ref{fig:hitomi_spec} for Suzaku/XIS and Hitomi/HXI, respectively.  The data in all altitude layers are well fitted with this model.  Tables~\ref{tab:NOcol_suzaku} and \ref{tab:NOcol_hitomi} show example N$+$O column densities measured with Suzaku/XIS and Hitomi/HXI, respectively.  In general, the column densities in different epochs are in good agreement.  We also note that results from Suzaku and Hitomi agree with each other in the altitude 90--100\,km.

It is well known that the O/N ratio, at least at the higher altitudes of our interest, changes with season, latitude, geomagnetic activity and so forth \cite<e.g.,>{2015E&SS....2....1M}.  Therefore, we took the N/O ratio from the NRLMSISE-00 model at the same date as the observation.  There are also diurnal variations of the O/N ratio.  According to the NRL model, the N/O ratio could vary in a day by 3\%, 15\%, and 30\% at altitudes 100\,km, 150\,km, and 200\,km, respectively.  To check this effect, we ran the spectral fitting with different relative abundances expected at different times.  As a result, the N$+$O column densities could change by at most 2\%, 10\%, and 20\% at altitudes 100\,km, 150\,km, and 200\,km, respectively.  These are less than the statistical uncertainties at the relevant altitudes.

\begin{figure}
\includegraphics[width=35pc,angle=0]{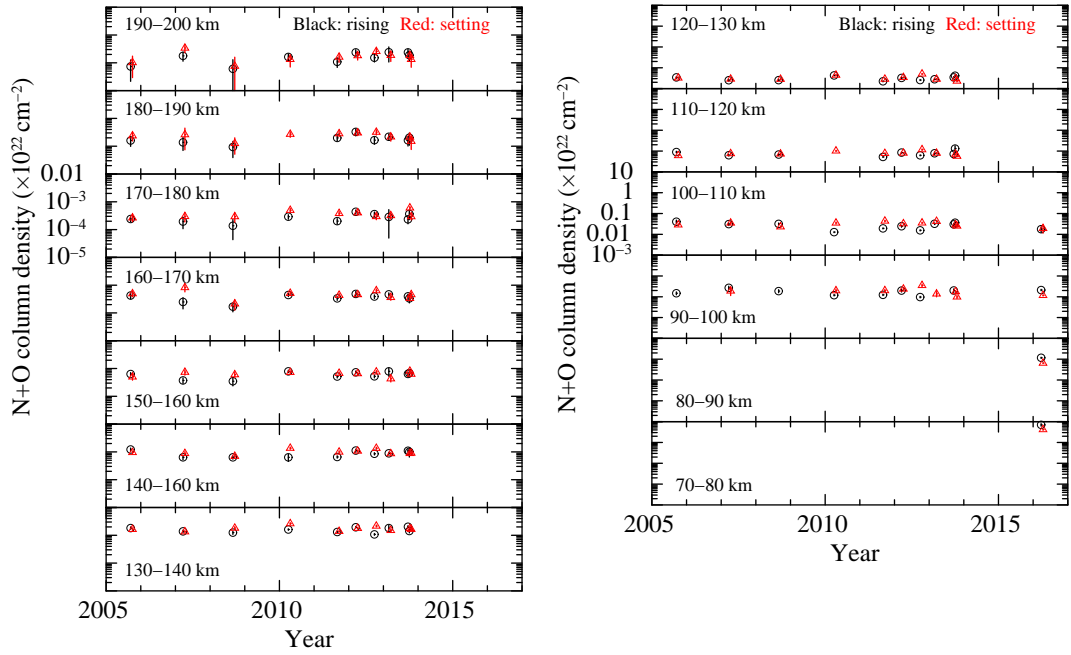}
\caption{N$+$O column densities as a function of time for each altitude layer as indicated in the upper left corner of each panel.  Black circles and red triangles are responsible for rising and setting in each observation epoch, respectively.  The time for the combined 100023010, 100023020, and 101011060 data is assumed to be that of 100023020.  For clarity, we shift the setting data by $+$20\,days.}
\label{fig:NOcol_trend}
\end{figure}

Figure~\ref{fig:NOcol_trend} shows all the measured N$+$O column densities as a function of time.  Each panel exhibits results for a different atmospheric layer.  No clear long-term/seasonal variations can be found at all layers.  To quantitatively estimate the long-term trend, we fitted the data with a linear function.  The best-fit time-variation rates are summarized in Table~\ref{tab:NOcol_trend}, from which we can conclude that the column densities stay constant within 10\% between 2005 and 2015.

In this context, we computed an average of the ten Suzaku observation epochs.  We also take standard deviations of the ten Suzaku data sets as our conservative measurement errors; the standard deviations are generally a few times larger than the statistical uncertainties on each data point for a single Obs.ID.  The results are listed in the right-end column of Table~\ref{tab:NOcol_suzaku}.  

\begin{figure}
\includegraphics[width=35pc,angle=0]{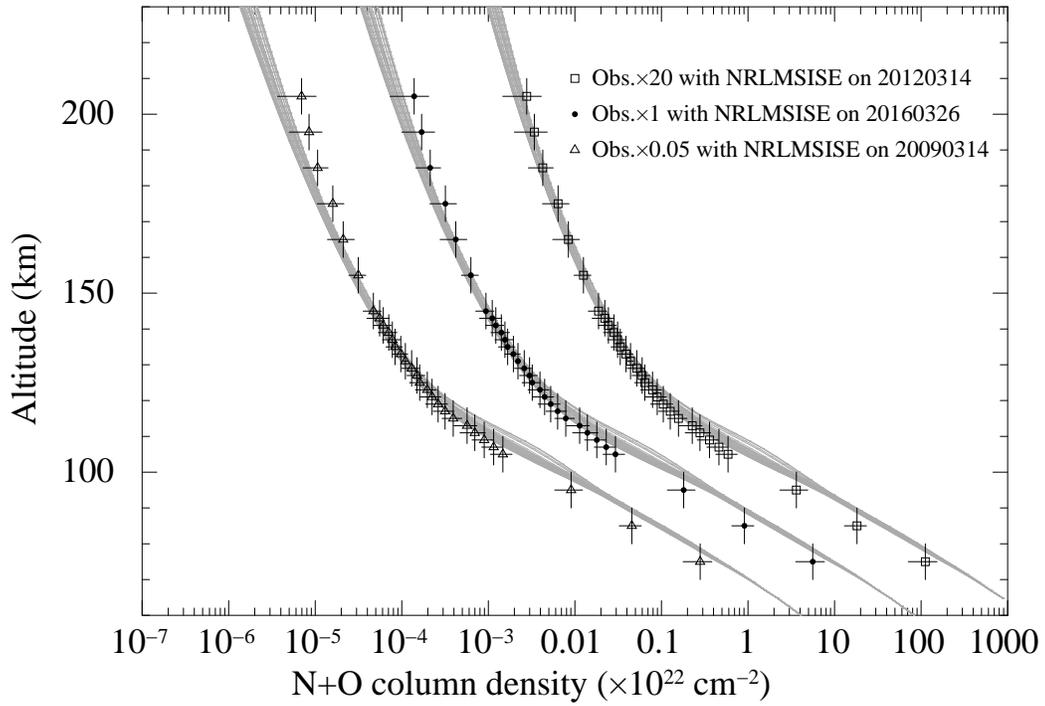}
\caption{Atmospheric N+O column densities as a function of altitude.  The bottom five data points are obtained with Hitomi/HXI, whereas other data are obtained with Suzaku/XIS, which are means and standard deviations of the ten data sets (Obs.IDs) listed in Table~\ref{tab:NOcol_suzaku}.  Gray lines represent predictions of the NRLMSISE-00 model at various hours of a day.  For clarity, we shift the data and model horizontally by multiplying factors of 20 (NRL model on 2012-03-14, the solar maximum) or 1/20 (NRL model on 2009-03-14, the solar minimum).}
\label{fig:NOcol}
\end{figure}

Figure~\ref{fig:NOcol} compares our measured N+O column densities with those expected by the NRLMSISE-00 model at three different dates: 1) 2009-03-14 for the solar minimum, 2) 2012-03-14 for the solar maximum, when Obs.ID~106014010 data were taken, 3) 2016-03-26 for the intermediate phase, when Obs.ID~100044010 data were taken.  The model curves are calculated at a location (latitude, longitude) = (0$^\circ$, 0$^\circ$) and times from 0:00 to 23:00 (UT) by a step of an hour to take account of the local time dependence.  In fact, the models at three epochs agree with each other below 150\,km altitude.  This is reasonable because the region below 150\,km altitude is almost independent on the solar cycle \cite<e.g.,>{2015E&SS....2....1M}.  Our measurements are in general agreement with the model.  

%We note in Figure~\ref{fig:NOcol} that errors below $\sim$100\,km altitude are systematically smaller than those in the upper altitudes.  This is because errors on the five data points from the bottom represent statistical uncertainties for the single Hitomi/HXI observation, whereas those on other data points represent standard deviations for ten Suzaku/XIS observations.  Without averaging, the Hitomi/HXI data could be biased.  However, fortunately, we expect that the ELV accuracy for Hitomi is significantly improved from Suzaku, because the possible major error source on ELV, i.e., the wrong coordinate system, suspected for Suzaku was resolved for Hitomi.  Nonetheless, we introduced $\pm\Delta$ELV2 of $\sim$0.069$^\circ$ to match the set and rise profiles for Hitomi (see Figure~\ref{fig:hitomi_lc}).  This is relatively small compared with those for Suzaku, but the presence of ELV shift indicates unresolved systematic errors on ELV even for Hitomi.

We then invert the tangential column number density to the local number density at the tangent point, using the technique developed by \citeA{1972P&SS...20.1727R}.  Specifically, we followed their technique to stabilize the small random errors in the data, by using the exponential form to approximate the atmospheric column density distribution.  In this procedure, we first need to obtain the normalization and decay constant of the exponential function at each data point.  To this end, the data point of our interest and its surrounding two data points in both sides are fitted with an exponential function.  Then, the best-fit parameters for the exponential function are used to derive the local density, by using Equation (9) in \citeA{1972P&SS...20.1727R}.  The errors are propagated according to the equation in the appendix of \citeA{1972P&SS...20.1727R}.  As a result, we obtain a vertical density distribution in Figure~\ref{fig:NOdens}.  We also plot a prediction by NRLMSISE-00 on 2016-03-26 when the Hitomi/HXI data were obtained.  As expected from Figure~\ref{fig:NOcol}, the density profile is in general agreement with the NRLMSISE-00 model.  However, we can see a significant density deficit at altitudes around 100\,km, which is magnified in the inset of Figure~\ref{fig:NOdens}.  

\section{Discussion}

\begin{figure}
\includegraphics[width=35pc,angle=0]{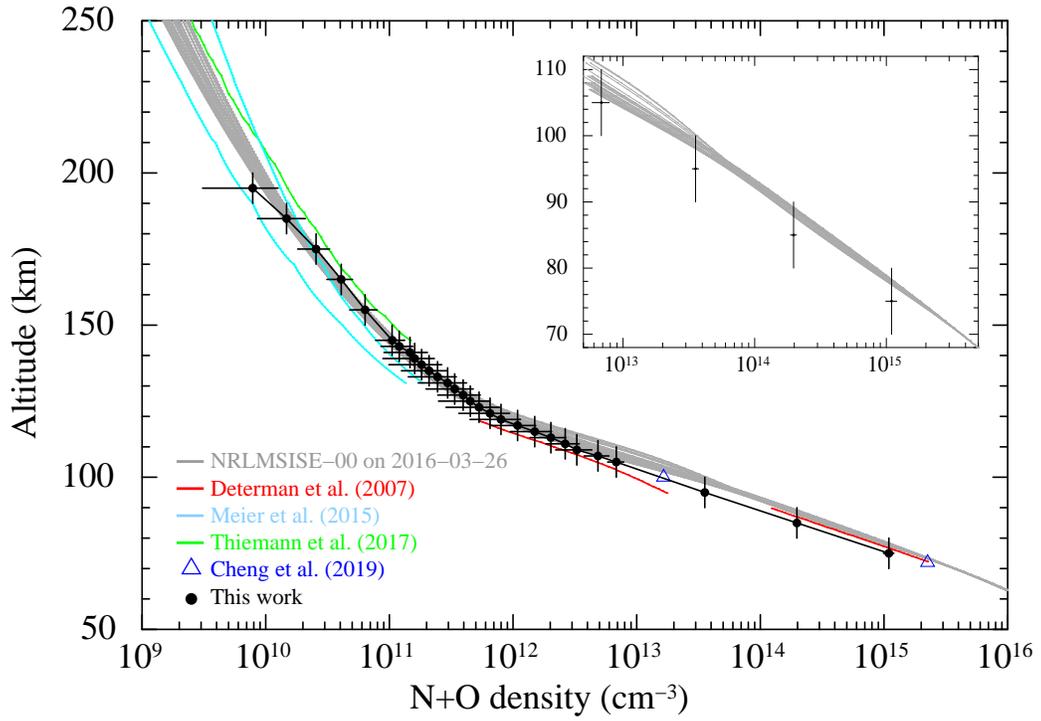}
\caption{Atmospheric N+O densities as a function of altitude, which is inverted from the column density in Figure~\ref{fig:NOcol}.  The gray lines represent the NRLMSISE-00 models at various hours of the day 2016-03-26.  Other data from recent literatures are also plotted.  The inset shows a close-up view of the lowest altitudes, where our density measurements are significantly smaller than the NRLMSISE-00 model.}
\label{fig:NOdens}
\end{figure}

We have measured the atmospheric N+O density in the altitude range of 70--200\,km and the latitude of $-$20$^\circ$--40$^\circ$ during a period from 2005 to 2016, with the X-ray astronomy satellites Suzaku and Hitomi.  We analyzed data from 219 occultation scans in total.  To improve the photon statistics and effective sampling rates for Suzaku, we accumulated all occultations in each Obs.ID or a few Obs.IDs.  Consequently, we derived 10 observation sequences for Suzaku.  Because we do not see systematic long-term trend for N+O column densities, we constructed a single, averaged vertical density profile as shown in Figure~\ref{fig:NOdens}.  For comparison, Figure~\ref{fig:NOdens} also shows recent measurements that overlap the altitude range of our interest \cite{2007JGRA..112.6323D,2015E&SS....2....1M,2017SpWea..15.1649T,2020Atmos..11..341C}.  We here focus on literature that lists N and O densities, excluding measurements of the O$_2$ density \cite <e.g.,>{2007JGRD..11216308L}, as O$_2$ is a minor species in the thermosphere.  When plotting the literature data in Figure~\ref{fig:NOdens}, we converted the data (either N$_2$+O or total mass density) into the total number densities of N and O, assuming the altitude-dependent atmospheric composition in the NRLMSISE-00 model.  

Our data are consistent with results from TIMED/SABER's infrared (IR) observations at 72\,km and 100\,km \cite{2020Atmos..11..341C}, as well as RXTE/PCA's X-ray occultation observations at altitudes 73--93\,km \cite{2007JGRA..112.6323D}.  Our data are marginally larger than the RXTE/PCA measurements in altitudes 100--120\,km.  It should be noted, however, that the measurement by \citeA{2007JGRA..112.6323D} itself seems to be subject to some systematic uncertainties, because their analysis, based on lower-energy emission (3--19\,keV and 2--12\,keV for PCA and USA, respectively) and a model with a single-density scalar at all altitudes, resulted in the density in altitudes 70--150\,km to be $\sim$1.7 times larger than that from the 3.0--5.4\,keV passband and SABER (which appears to be adopted as their final result).  Their higher-density result is fully consistent with our measurement, and in fact better connects to the density in the lower altitude \cite{2007JGRA..112.6323D}.  Therefore, it is unclear whether our results are truely different from those by \citeA{2007JGRA..112.6323D} at altitudes 100--120\,km.  Above 120\,km, our data generally agree with two previous measurements by TIMED/GUVI \cite<>[data taken on day 77 in 2002]{2015E&SS....2....1M} and PROBA2/LYRA \cite{2017SpWea..15.1649T}.   It should be noted that the LYRA data below 200\,km could be overestimated by 5--7\% \cite{2017SpWea..15.1649T}.  If this systematic bias is corrected, then the LYRA data will be in even better agreement with ours.  However, the GUVI data taken on day 194 in 2006 \cite{2015E&SS....2....1M} show a significantly lower density than ours.  This might be partly due to the latitude dependence on the atomic O density, which decreases more rapidly with increasing height at higher latitudes.  We conclude that our data are in reasonable agreement with earlier density measurements.  

%Thiemann+2017 --> 2010-12-07 + 2013-12-07
%Determan+2007 --> 2005-11-14 (PCA), USA unknown 
%Meier+2015 --> 2002-03-18 + 2006-07-13
%Cheng+2020 --> 2017-09-08 + 2017-05-09

\begin{figure}
\includegraphics[width=35pc,angle=0]{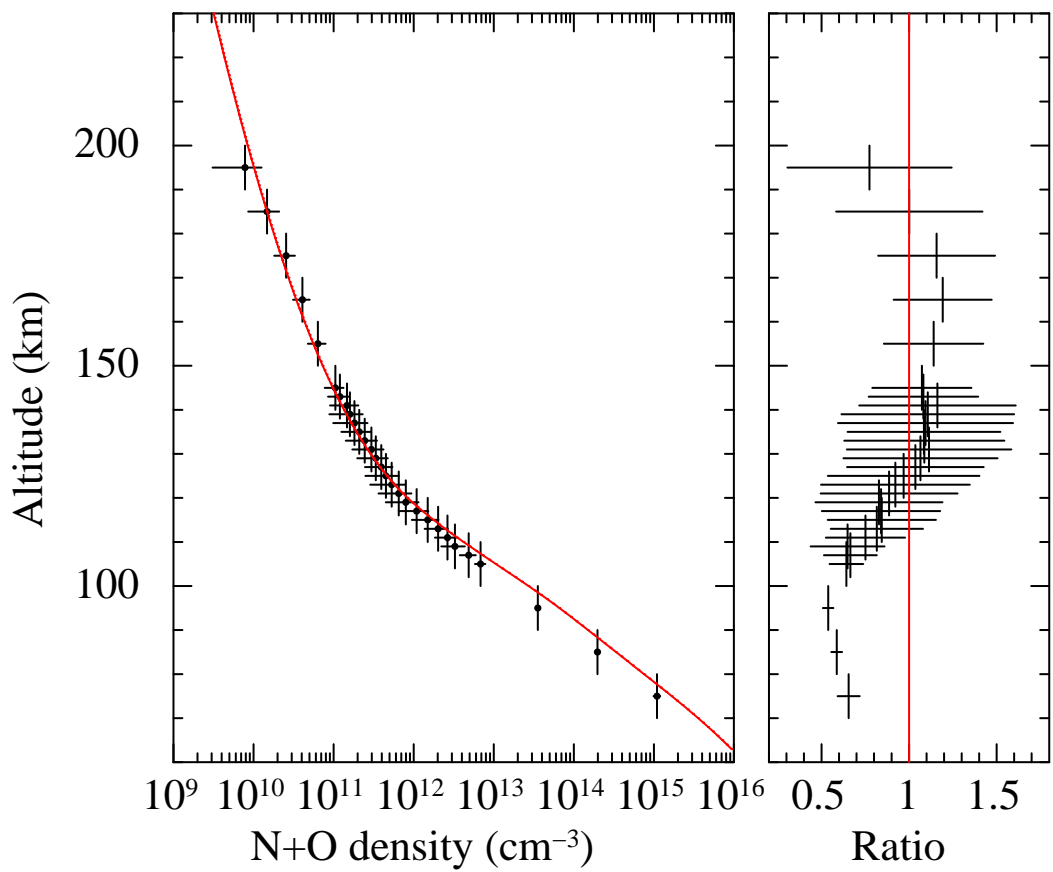}
\caption{Left: Same as Figure~\ref{fig:NOdens} but the model curve in red is the mean on 2016-03-26.  Right: Ratio between the data and model shown in the left panel.}
\label{fig:ratio}
\end{figure}

To compare our measurement and the model more quantitatively, we plot in Figure~\ref{fig:ratio} the data with a time-averaged density model on 2016-03-26, when Hitomi/HXI data were taken.  The right-hand panel shows the data-to-model ratio as a function of altitude.  As mentioned in the previous section, the measured density profile is in good agreement with the NRLMSISE-00 model, except for the altitude range of 70--110\,km, where the data are significantly smaller than the model with a maximum deficit of $-50$\% at altitude of $\sim$95\,km.  

Both theoretical models and observations have suggested that the increasing greenhouse gas (e.g., CO$_2$ and CH$_4$) concentration in the troposphere causes the upper atmosphere to cool and contract, resulting in a corresponding density decrease at a given height \cite<e.g.,>{1989GeoRL..16.1441R,1998AnGeo..16.1501A,2000GeoRL..27.1523K}.  In the troposphere, CO$_2$ is optically thick and traps IR radiation emitted by the Earth's surface.  In the stratosphere and above, CO$_2$ is optically thin and emits infrared radiation to space, which cools and contracts these regions.  \citeA{2015AdSpR..56..773E} summarized long-term density trends from both modeling studies and observations.  Although there is a significant scatter among the data, the trend is all negative (decreasing density) in the thermosphere.  It is interesting to note that \citeA{2006JASTP..68.1879A} predicted that a layer near 110\,km shows a maximum density reduction of $-6.5$\% per decade, which was later quantitatively confirmed by radar observations of meteor trails \cite{2014GeoRL..41.6919S}.  The ratio between our data and the NRLMSISE-00 model in Figure~\ref{fig:ratio} shows a maximum of the density decline at $\sim$100\,km altitude, which is qualitatively consistent with the trend found by \citeA{2006JASTP..68.1879A}.  Quantitatively, if the data used to construct the NRL model was taken half century ago, it is reasonable to expect a density reduction of $-$33\% between our data and the NLR model.  This is close to the discrepancy between our measurement and the model.  However, we cannot yet rule out a possibility that the NRL model overestimated the density at altitudes around 100\,km.  In fact, a new model of MSIS, i.e., NRLMSIS 2.0 \cite{Emmert2020}, was released during the preparation of this paper.  By using the python package {\tt pymsis} (available at {\tt https://pypi.org/project/pymsis/}), we calculated updated density profiles expected on 2016-03-26, and found that the new model gives 20--30\% lower densities at altitudes 70--110\,km.  Therefore, the new model shortens the gap between the data and the model.  Still, our data are slightly lower than the new model at the altitude of $\sim$100\,km.

No significant long-term density variation was found within our data.  Also, there is no clear density difference compared with past observations.  One caution in discussing density variations would be the fact that the N$+$O column densities (or half-flux altitudes) show statistically significant differences from observation to observation, as well as between settings and risings.  As far as we investigated, there seem no instrumental systematic uncertainties to explain this variation.  However, there may be missing errors.  Therefore, in the future work, it is important to carefully reexamine possible systematic uncertainties to discuss temporal variations of the atmospheric density.

\begin{figure}
\includegraphics[width=35pc,angle=0]{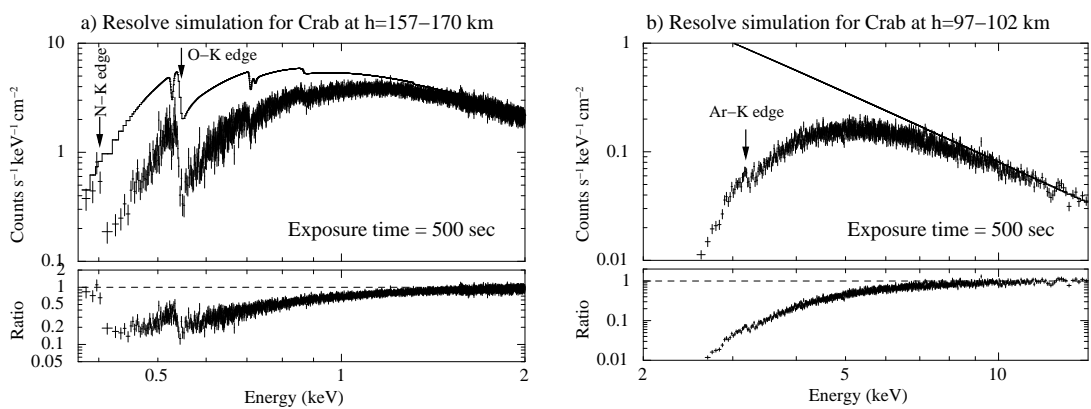}
\caption{Crab Nebula's X-ray spectra to be obtained with XRISM/Resolve, at altitude ranges 157--170\,km (left) and 97--102\,km (right).  Lower panels show ratios between the data and the model which does not take account of the Earth's atmospheric absorption.  Due to the atmospheric absorption, the K-shell edges of N, O, and Ar are clearly seen, from which we will be able to measure the chemical composition of the atmosphere.}  
\label{fig:resolve_sim}
\end{figure}

Occultations of X-ray astronomical sources have been detected with other X-ray astronomy satellites in low Earth orbits, including terminated ones such as Ginga \cite{1987ApL....25..223M} and ASCA \cite{1994PASJ...46L..37T}, as well as in-operation ones such as NuSTAR \cite{2013ApJ...770..103H} and NICER \cite{2016SPIE.9905E..1HG}.  Analyses of the data from these satellites will allow us to investigate a long-term trend of the atmospheric density.  Specifically, NuSTAR is suitable to diagnose the density deficit region at the altitude $\sim$100\,km, thanks to its good sensitivity in a wide energy range of 3--80\,keV.  NICER has an unprecedented throughput that will allow us to obtain the density profile every single occultation.  Also, the X-Ray Imaging and Spectroscopy Mission \cite<XRISM:>{2018SPIE10699E..22T}, the Japan-US X-ray astronomy mission scheduled to be launched in 2022, will carry an X-ray micro-calorimeter \cite<Resolve:>{2018JLTP..193..991I} that will allow for high-resolution spectroscopy of a resolution of $\Delta$$E$ $\sim$5\,keV with little energy dependence in 0.2--12\,keV.  To demonstrate its capability, we simulate XRISM/Resolve spectra expected during the occultation of the Crab Nebula, as shown in Figure~\ref{fig:resolve_sim} left and right for altitude ranges 157--170\,km and 97--102\,km, respectively.  Absorption edges of N at 0.41\,keV and O at 0.54\,keV can be seen at higher tangential altitudes, and that of Ar at 3.2\,keV can be seen at lower tangential altitudes.  The depths of these edges will tell us the composition of the atmosphere.  These simulations are performed for an exposure time of 500\,s.  This exposure time will be accumulated by $\sim$100 and $\sim$300 occultations for altitude ranges 157--170\,km and 97--102\,km, respectively.  Therefore, it will take a few years for us to take the spectrum of this quality from calibration data alone, but it is certainly doable with XRISM/Resolve.  

Finally, we point out that measuring atmospheric densities with X-ray astronomy satellites has just started, and thus more experiences are essential to obtain deeper insights into unresolved issues such as the potential systematic uncertainties on the tangential altitude.  

\section{Conclusions}

By analyzing data during atmospheric occultations of the Crab Nebula, obtained with two X-ray astronomy satellites Suzaku and Hitomi, we measured a vertical density profile of the Earth's atmosphere in altitudes of 70--200\,km.  We provided one density profile, by averaging 219 occultations taken during 2005--2016.  The vertical density profile is generally consistent with a prediction by the empirical NRLMSISE-00 model.  Our measurement is also generally consistent with several earlier measurements \cite{2007JGRA..112.6323D,2015E&SS....2....1M,2017SpWea..15.1649T,2020Atmos..11..341C}.  We found a significant density deficit at the altitude range of 70--110\,km.  The strong density deficit at $\sim$100\,km altitude is qualitatively consistent with a model prediction and observation that claims long-term cooling in the upper atmosphere due to IR radiation from increasing greenhouse gases \cite{2006JASTP..68.1879A,2014GeoRL..41.6919S}.  It is important to monitor the atmospheric density to clarify if the possible long-term trend is true or not.  This work can be done by analyses of data acquired with past, in-operation, and future X-ray astronomy satellites.  In addition, with the upcoming XRISM, we will be able to measure the composition of the upper atmosphere from K-shell absorption edges of N, O, and Ar.  

\acknowledgments
All the data used in this paper can be found at NASA's HEASARC website, {\tt https://heasarc.gsfc.nasa.gov}.  We are grateful to all the members of Suzaku and Hitomi teams.  We thank Ms.\ Mina Ogawa and Dr.\ Tadayasu Dotani for providing us with information about possible Suzaku's position uncertainty.  We also thank anonymous referees for a number of constructive comments that significantly improved the quality of the original manuscript.  This work was supported by the Japan Society for the Promotion of Science KAKENHI grant numbers 20K20935 (SK and MST), 16H03983 (KM).  This work was partly supported by Leading Initiative for Excellent Young Researchers, MEXT, Japan.

%\bibliography{reference}

\clearpage

%  Numbered lines in equations:
%  To add line numbers to lines in equations,
%  \begin{linenomath*}
%  \begin{equation}
%  \end{equation}
%  \end{linenomath*}

\begin{longtable}[c]{lcccc}
 \caption{Summary of Crab occultations analyzed in this paper.}
 \label{tab:data}
 \\
 %------ 最初のページの表の最上部 ----
 \hline 
Instrument & UT at h = 100\,km$^a$ & Tangent point$^b$ & Local time$^c$ & Occultation type (\#) \\
(Obs.ID) & (HH:MM:SS) & Lat, Long ($^\circ$) & (HH:MM:SS) & \\
\hline
 \endfirsthead
 %------ 2ページ以降の表の最上部 ----
 \multicolumn{5}{l}{\small }\\
 \hline
Instrument & UT at h = 100\,km$^a$ & Tangent point$^b$ & Local time$^c$ & Occultation type (\#)\\ 
(Obs.ID) & (HH:MM:SS) & Lat, Long ($^\circ$) & (HH:MM:SS) & \\
 \hline
 \endhead
 %----- ページの表の最下部 --------
 \hline
\multicolumn{5}{l}{\small $^a$Universal time when the line-of-sight tangential altitude ($h$) becomes 100\,km.}\\
\multicolumn{5}{l}{\small $^b$Latitude and longitude of the tangent point at $h = 100$\,km.}\\
\multicolumn{5}{l}{\small $^c$Local time of the tangent point at $h = 100$\,km.}\\
 \endfoot 
 %----- 最終ページの表の最下部 --------
 \hline
%\multicolumn{5}{l}{\small\it End of table}\\  
\multicolumn{5}{l}{\small $^a$Universal time when the line-of-sight tangential altitude ($h$) becomes 100\,km.}\\  
\multicolumn{5}{l}{\small $^b$Latitude and longitude of the tangent point at $h = 100$\,km.}\\  
\multicolumn{5}{l}{\small $^c$Local time of the tangent point at $h = 100$\,km.}\\
 \endlastfoot
 %----------------------------------------------------------------
\multirow{5}{*}{
\begin{tabular}{l}
Suzaku/XIS\\(100023010)
\end{tabular}}
 & 2005-09-15, 14:11:32 & 20.06, 138.19 & 23:24:24 & Rising (1)\\
 & 2005-09-15, 15:47:28 & 20.23, 114.02 & 23:23:23 & Rising (2)\\
 & 2005-09-15, 17:23:24 & 20.39, 89.85 & 23:22:22 &Rising (3)\\
 & 2005-09-15, 18:22:36 & -20.45, 255.62 & 11:25:25 &Setting (4)\\
 & 2005-09-15, 18:59:21 & 20.59, 65.74 & 23:22:22 &Rising (5)\\
\hline
\multirow{8}{*}{
\begin{tabular}{l}
Suzaku/XIS\\(100023020)
\end{tabular}}
 & 2005-09-15, 19:58:33 & -20.55, 231.59 & 11:24:24 &Setting (6)\\
 & 2005-09-15, 20:35:17 & 20.76, 41.57 & 23:21:21 &Rising  (7)\\
 & 2005-09-15, 21:34:29 & -20.63, 207.50 & 11:24:24 &Setting  (8)\\
 & 2005-09-15, 22:11:13 & 20.93, 17.41 & 23:20:20 &Rising  (9)\\
 & 2005-09-15, 23:10:25 & -20.71, 183.41 & 11:24:24 &Setting  (10)\\
 & 2005-09-15, 23:47:09 & 21.09, 353.25 & 23:20:20 &Rising  (11)\\
 & 2005-09-16, 00:46:21 & -20.79, 159.33 & 11:23:23 &Setting  (12)\\
 & 2005-09-16, 01:23:06 & 21.29, 329.14 & 23:19:19 &Rising  (13)\\
\hline
\multirow{2}{*}{
\begin{tabular}{l}
Suzaku/XIS\\(101011060)
\end{tabular}}
 & 2006-09-18, 21:34:10 & -3.99, 35.19 & 23:54:54 &Rising (14)\\
 & 2006-09-18, 22:38:13 & 39.57, 220.10 & 13:18:18 &Setting  (15)\\
\hline
\multirow{22}{*}{
\begin{tabular}{l}
Suzaku/XIS\\(102019010) 
\end{tabular}}
 & 2007-03-20, 11:03:59 & 26.72, 202.33 & 00:33:33 &Setting (16)\\
 & 2007-03-20, 18:04:12 & -22.64, 273.10 & 12:16:16 &Rising  (17)\\
 & 2007-03-20, 19:03:27 & 25.99, 81.59 & 00:29:29 &Setting (18)\\
 & 2007-03-20, 19:40:06 & -22.59, 249.06 & 12:16:16 &Rising (19)\\
 & 2007-03-20, 20:39:21 & 25.83, 57.46 & 00:29:29 &Setting (20)\\
 & 2007-03-20, 21:16:00 & -22.54, 225.01 & 12:16:16 &Rising  (21)\\
 & 2007-03-20, 22:15:15 & 25.67, 33.33 & 00:28:28 &Setting  (22)\\
 & 2007-03-20, 22:51:54 & -22.50, 200.95 & 12:15:15 &Rising  (23)\\
 & 2007-03-20, 23:51:09 & 25.51, 9.20 & 00:27:27 &Setting  (24)\\
 & 2007-03-21, 00:27:48 & -22.45, 176.90 & 12:15:15 &Rising  (25)\\
 & 2007-03-21, 01:27:02 & 25.38, -14.98 & 00:27:27 &Setting  (26)\\
 & 2007-03-21, 02:03:42 & -22.40, 152.84 & 12:15:15 &Rising  (27)\\
 & 2007-03-21, 03:02:56 & 25.22, 320.88 & 00:26:26 &Setting  (28)\\
 & 2007-03-21, 03:39:36 & -22.35, 128.79 & 12:14:14 &Rising  (29)\\
 & 2007-03-21, 04:38:50 & 25.06, 296.75 & 00:25:25 &Setting  (30)\\
 & 2007-03-21, 05:15:30 & -22.30, 104.73 & 12:14:14 &Rising  (31)\\
 & 2007-03-21, 06:14:44 & 24.90, 272.62 & 0:25:25 &Setting  (32)\\
 & 2007-03-21, 06:51:24 & -22.25, 80.67 & 12:14:14 &Rising  (33)\\
 & 2007-03-21, 07:50:37 & 24.76, 248.44 & 00:24:24 &Setting  (34)\\
 & 2007-03-21, 08:27:18 & -22.20, 56.62 & 12:13:13 &Rising  (35)\\
 & 2007-03-21, 09:26:31 & 24.60, 224.31 & 00:23:23 &Setting  (36)\\
 & 2007-03-21, 10:03:12 & -22.14, 32.56 & 12:13:13 &Rising  (37)\\
\hline
\multirow{20}{*}{
\begin{tabular}{l}
Suzaku/XIS\\(103007010)
\end{tabular}}
 & 2008-08-27, 08:53:46 & 8.65, 242.05 & 01:01:01 &Rising (38)\\
 & 2008-08-27, 10:29:39 & 8.85, 217.91 & 01:01:01 &Rising  (39)\\
 & 2008-08-27, 12:05:33 & 9.07, 193.82 & 01:00:00 & Rising  (40)\\
 & 2008-08-27, 13:04:46 & -13.25, -4.33 & 12:47:47 &Setting  (41)\\
 & 2008-08-27, 13:41:26 & 9.27, 169.68 & 01:00:00 &Rising  (42)\\
 & 2008-08-27, 15:17:19 & 9.46, 145.53 & 00:59:59 &Rising  (43)\\
 & 2008-08-27, 16:53:12 & 9.66, 121.39 & 00:58:58 &Rising  (44)\\
 & 2008-08-27, 18:29:06 & 9.88, 97.30 & 00:58:58 &Rising  (45)\\
 & 2008-08-27, 19:28:19 & -13.86, 259.36 & 12:45:45 &Setting  (46)\\
 & 2008-08-27, 20:04:59 & 10.08, 73.16 & 00:57:57 &Rising  (47)\\
 & 2008-08-27, 21:04:12 & -14.00, 235.27 & 12:45:45 &Setting  (48)\\
 & 2008-08-27, 21:40:52 & 10.27, 49.02 & 00:56:56 &Rising  (49)\\
 & 2008-08-27, 22:40:05 & -14.15, 211.18 & 12:44:44 &Setting  (50)\\
 & 2008-08-28, 00:15:58 & -14.30, 187.09 & 12:44:44 &Setting  (51)\\
 & 2008-08-28, 00:52:38 & 10.66, 0.74 & 00:55:55 &Rising  (52)\\
 & 2008-08-28, 01:51:51 & -14.44, 163.01 & 12:43:43 &Setting  (53)\\
 & 2008-08-28, 02:28:32 & 10.89, 336.66 & 00:55:55 &Rising  (54)\\
 & 2008-08-28, 03:27:44 & -14.58, 138.93 & 12:43:43 &Setting  (55)\\
 & 2008-08-28, 04:04:25 & 11.09, 312.52 & 00:54:54 &Rising  (56)\\
 & 2008-08-28, 05:03:37 & -14.72, 114.84 & 12:42:42 &Setting  (57)\\
\hline
\multirow{21}{*}{
\begin{tabular}{l}
Suzaku/XIS\\(105002010)
\end{tabular}}
 & 2010-04-05, 14:05:54 & -6.79, 124.97 & 22:25:25 &Setting  (58)\\
 & 2010-04-05, 15:41:49 & -6.51, 101.04 & 22:25:25 &Setting  (59)\\
 & 2010-04-05, 17:17:43 & -6.27, 77.05 & 22:25:25 &Setting  (60)\\
 & 2010-04-05, 18:53:38 & -5.99, 53.11 & 22:26:26 &Setting  (61)\\
 & 2010-04-05, 19:25:54 & 39.68, 204.51 & 09:03:03 &Rising  (62)\\
 & 2010-04-05, 20:29:33 & -5.71, 29.17 & 22:26:26 &Setting  (63)\\
 & 2010-04-05, 21:01:46 & 39.67, 180.51 & 09:03:03 &Rising  (64)\\
 & 2010-04-05, 22:05:28 & -5.43, 5.23 & 22:26:26 &Setting  (65)\\
 & 2010-04-05, 22:37:38 & 39.65, 156.52 & 09:03:03 &Rising  (66)\\
 & 2010-04-06, 00:13:29 & 39.64, 132.46 & 09:03:03 &Rising  (67)\\
 & 2010-04-06, 01:49:21 & 39.62, 108.47 & 09:03:03 &Rising  (68)\\
 & 2010-04-06, 03:25:13 & 39.60, 84.48 &09:03:03 & Rising  (69)\\
 & 2010-04-06, 05:01:05 & 39.58, 60.49 &09:03:03 & Rising  (70)\\
 & 2010-04-06, 06:05:02 & -4.04, 245.48  & 22:26:26 &Setting  (71)\\
 & 2010-04-06, 06:36:57 & 39.55, 36.50 & 09:02:02 & Rising  (72)\\
 & 2010-04-06, 07:40:57 & -3.75, 221.53 & 22:27:27 &Setting  (73)\\
 & 2010-04-06, 08:12:48 & 39.53, 12.44 & 09:02:02 & Rising  (74)\\
 & 2010-04-06, 09:16:52 & -3.46, 197.58 & 22:27:27 &Setting  (75)\\
 & 2010-04-06, 09:48:40 & 39.50, 348.45 & 09:02:02 &Rising  (76)\\
 & 2010-04-06, 10:52:47 & -3.18, 173.63 & 22:27:27 &Setting  (77)\\
 & 2010-04-06, 11:24:32 & 39.47, 324.45 & 09:02:02 &Rising  (78)\\
\hline
\multirow{27}{*}{
\begin{tabular}{l}
Suzaku/XIS\\(106012010)
\end{tabular}}
 & 2011-09-01, 06:39:30 & 37.96, 252.11 & 23:27:27 &Rising  (79)\\
 & 2011-09-01, 07:45:03 & 5.60, 78.42 & 12:58:58 &Setting  (80)\\
 & 2011-09-01, 08:15:20 & 37.87, 228.21 & 23:28:28 &Rising (81)\\
 & 2011-09-01, 09:20:56 & 5.95, 54.53 & 12:59:59 &Setting (82)\\
 & 2011-09-01, 09:51:09 & 37.78, 204.24 & 23:28:28 &Rising (83)\\ 
 & 2011-09-01, 10:56:49 & 6.30, 30.64 & 12:59:59 &Setting (84)\\
 & 2011-09-01, 11:26:59 & 37.69, 180.34 & 23:28:28 &Rising (85)\\
 & 2011-09-01, 12:32:42 & 6.65, 6.76 & 12:59:59 &Setting (86)\\
 & 2011-09-01, 13:02:49 & 37.59, 156.43 & 23:28:28 &Rising(87)\\ 
 & 2011-09-01, 14:38:39 & 37.49, 132.53 & 23:28:28 &Rising (88)\\
 & 2011-09-01, 16:14:28 & 37.40, 108.56 & 23:28:28 &Rising (89)\\
 & 2011-09-01, 17:50:18 & 37.29, 84.66 & 23:28:28 &Rising (90)\\
 & 2011-09-01, 18:56:13 & 8.04, 271.18 & 13:00:00 &Setting (91)\\
 & 2011-09-01, 19:26:08 & 37.18, 60.76 & 23:29:29 &Rising (92)\\
 & 2011-09-01, 20:32:07 & 8.42, 247.35 & 13:01:01 &Setting (93)\\
 & 2011-09-01, 21:01:58 & 37.07, 36.85 & 23:29:29 &Rising (94)\\
 & 2011-09-01, 22:08:00 & 8.77, 223.46 & 13:01:01 &Setting (95)\\
 & 2011-09-01, 22:37:48 & 36.96, 372.94 & 23:29:29 &Rising (96)\\
 & 2011-09-01, 23:43:53 & 9.12, 199.57 & 13:02:02 &Setting (97)\\
 & 2011-09-02, 00:13:38 & 36.84, 349.02 & 23:29:29 &Rising (98)\\
 & 2011-09-02, 01:19:46 & 9.47, 175.68 & 13:02:02 & Setting (99)\\
 & 2011-09-02, 01:49:28 & 36.73, 325.11 & 23:29:29 &Rising (100)\\
 & 2011-09-02, 02:55:39 & 9.83, 151.80 & 13:02:02 &Setting (101)\\
 & 2011-09-02, 03:25:18 & 36.61, 301.20 & 23:30:30 &Rising (102)\\
 & 2011-09-02, 04:31:32 & 10.18, 127.91 & 13:03:03 &Setting (103)\\
 & 2011-09-02, 05:01:08 & 36.49, 277.29 & 23:30:30 &Rising (104)\\
 & 2011-09-02, 06:07:25 & 10.53, 104.03 & 13:03:03 &Setting (105)\\
\hline
\multirow{28}{*}{
\begin{tabular}{l}
Suzaku/XIS\\(106014010)
\end{tabular}}
 & 2012-03-14, 01:27:52 & 20.18, -12.20 & 00:39:39 &Setting  (106)\\
 & 2012-03-14, 02:04:37 & -20.13, 158.62 & 12:39:39 &Rising  (107)\\
 & 2012-03-14, 03:03:35 & 20.00, 323.71 & 00:38:38 &Setting  (108)\\
 & 2012-03-14, 03:40:20 & -20.04, 134.60 & 12:38:38 &Rising  (109)\\
 & 2012-03-14, 04:39:18 & 19.83, 299.61 & 00:37:37 &Setting  (110)\\
 & 2012-03-14, 05:16:03 & -19.95, 110.57 & 12:38:38 &Rising  (111)\\
 & 2012-03-14, 06:15:01 & 19.65, 275.52 & 00:37:37 &Setting  (112)\\
 & 2012-03-14, 06:51:46 & -19.86, 86.55 & 12:37:37 &Rising  (113)\\
 & 2012-03-14, 07:50:44 & 19.47, 251.43 & 00:36:36 &Setting  (114)\\
 & 2012-03-14, 08:27:29 & -19.76, 62.53 & 12:37:37 &Rising  (115)\\
 & 2012-03-14, 09:26:27 & 19.29, 227.33 & 00:35:35 &Setting  (116)\\
 & 2012-03-14, 11:02:10 & 19.11, 203.23 & 00:35:35 &Setting  (117)\\
 & 2012-03-14, 12:37:53 & 18.93, 179.13 & 00:34:34 &Setting  (118)\\
 & 2012-03-14, 14:13:36 & 18.75, 155.03 & 00:33:33 &Setting  (119)\\
 & 2012-03-14, 15:49:19 & 18.57, 130.93 & 00:33:33 &Setting  (120)\\
 & 2012-03-14, 17:25:02 & 18.38, 106.84 & 00:32:32 &Setting  (121)\\
 & 2012-03-14, 18:01:47 & -19.18, 278.38 & 12:35:35 &Rising  (122)\\
 & 2012-03-14, 19:00:45 & 18.20, 82.74 & 00:31:31 &Setting  (123)\\
 & 2012-03-14, 19:37:30 & -19.07, 254.36 & 12:34:34 &Rising  (124)\\
 & 2012-03-14, 20:36:28 & 18.02, 58.65 & 00:31:31 &Setting  (125)\\
 & 2012-03-14, 21:13:13 & -18.96, 230.33 & 12:34:34 &Rising  (126)\\
 & 2012-03-14, 22:12:11 & 17.83, 34.55 & 00:30:30 &Setting  (127)\\
 & 2012-03-14, 22:48:56 & -18.86, 206.30 & 12:34:34 &Rising  (128)\\
 & 2012-03-14, 23:47:54 & 17.65, 10.45 & 00:29:29 &Setting  (129)\\
 & 2012-03-15, 00:24:39 & -18.76, 182.27 & 12:33:33 &Rising  (130)\\
 & 2012-03-15, 01:23:37 & 17.47, -13.65 & 00:29:29 &Setting  (131)\\
 & 2012-03-15, 02:00:22 & -18.65, 158.24 & 12:33:33 &Rising  (132)\\
 & 2012-03-15, 02:59:20 & 17.29, 322.25 & 00:28:28 &Setting  (133)\\
\hline
\multirow{25}{*}{
\begin{tabular}{l}
Suzaku/XIS\\(107011010) 
\end{tabular}}
 & 2012-09-26, 06:04:30 & -15.67, 70.73 & 10:47:47 &Setting  (134)\\
 & 2012-09-26, 06:38:47 & 39.03, 226.21 & 21:43:43 &Rising  (135)\\
 & 2012-09-26, 07:40:11 & -15.51, 46.77 & 10:47:47 &Setting  (136)\\
 & 2012-09-26, 08:14:26 & 39.06, 202.16 & 21:43:43 &Rising  (137)\\
 & 2012-09-26, 09:50:06 & 39.11, 178.19 & 21:42:42 &Rising  (138)\\
 & 2012-09-26, 11:25:45 & 39.14, 154.14 & 21:42:42 &Rising  (139)\\
 & 2012-09-26, 13:01:24 & 39.18, 130.10 & 21:41:41 &Rising  (140)\\
 & 2012-09-26, 14:37:04 & 39.22, 106.13 & 21:41:41 &Rising  (141)\\
 & 2012-09-26, 16:12:43 & 39.25, 82.09 & 21:41:41 &Rising  (142)\\
 & 2012-09-26, 17:14:21 & -14.44, 263.18 & 10:47:47 &Setting  (143)\\
 & 2012-09-26, 17:48:22 & 39.28, 58.05 & 21:40:40 &Rising  (144)\\
 & 2012-09-26, 18:50:02 & -14.27, 239.22 & 10:46:46 &Setting  (145)\\
 & 2012-09-26, 19:24:02 & 39.32, 34.08 & 21:40:40 &Rising  (146)\\
 & 2012-09-26, 20:25:44 & -14.08, 215.30 & 10:46:46 &Setting  (147)\\
 & 2012-09-26, 20:59:41 & 39.34, 370.05 & 21:39:39 &Rising  (148)\\
 & 2012-09-26, 22:01:26 & -13.88, 191.38 & 10:46:46 &Setting  (149)\\
 & 2012-09-26, 22:35:21 & 39.37, 346.07 & 21:39:39 &Rising  (150)\\
 & 2012-09-26, 23:37:08 & -13.69, 167.46 & 10:46:46 &Setting  (151)\\
 & 2012-09-27, 00:11:00 & 39.40, 322.03 & 21:39:39 &Rising  (152)\\
 & 2012-09-27, 01:12:50 & -13.49, 143.54 & 10:46:46 &Setting  (153)\\
 & 2012-09-27, 01:46:39 & 39.43, 297.99 & 21:38:38 &Rising  (154)\\
 & 2012-09-27, 02:48:32 & -13.29, 119.62 & 10:47:47 &Setting  (155)\\
 & 2012-09-27, 03:22:19 & 39.46, 274.02 & 21:38:38 &Rising  (156)\\
 & 2012-09-27, 04:24:13 & -13.12, 95.64 & 10:46:46 &Setting  (157)\\
 & 2012-09-27, 04:57:58 & 39.48, 249.98 & 21:37:37 &Rising  (158)\\
\hline
\multirow{22}{*}{
\begin{tabular}{l}
Suzaku/XIS\\(107012010)
\end{tabular}}
 & 2013-02-27, 00:32:33 & -11.13, 3.30 & 00:45:45 &Setting (159)\\
 & 2013-02-27, 01:09:15 & 6.83, 178.21 & 13:02:02 &Rising (160)\\
 & 2013-02-27, 02:44:52 & 7.04, 154.16 & 13:01:01 &Rising (161)\\
 & 2013-02-27, 04:20:29 & 7.24, 130.10 & 13:00:00 &Rising (162)\\
 & 2013-02-27, 05:56:06 & 7.45, 106.05 & 13:00:00 &Rising (163)\\
 & 2013-02-27, 07:31:43 & 7.66, 81.99 & 12:59:59 &Rising (164)\\
 & 2013-02-27, 08:30:37 & -11.94, 243.17 & 00:43:43 &Setting (165)\\
 & 2013-02-27, 09:07:20 & 7.87, 57.93 & 12:59:59 &Rising (166)\\
 & 2013-02-27, 10:06:14 & -12.11, 219.16 & 00:42:42 &Setting (167)\\
 & 2013-02-27, 11:41:51 & -12.28, 195.15 & 00:42:42 &Setting (168)\\
 & 2013-02-27, 13:17:28 & -12.44, 171.15 & 00:42:42 &Setting (169)\\
 & 2013-02-27, 13:54:11 & 8.49, 345.78 & 12:57:57 &Rising (170)\\
 & 2013-02-27, 14:53:05 & -12.60, 147.15 & 00:41:41 &Setting (171)\\
 & 2013-02-27, 15:29:48 & 8.70, 321.73 & 12:56:56 &Rising (172)\\
 & 2013-02-27, 16:28:41 & -12.74, 123.08 & 00:41:41 &Setting (173)\\
 & 2013-02-27, 17:05:25 & 8.91, 297.68 & 12:56:56 &Rising (174)\\
 & 2013-02-27, 18:04:18 & -12.90, 99.08 & 00:40:40 &Setting (175)\\
 & 2013-02-27, 18:41:02 & 9.12, 273.62 & 12:55:55 &Rising (176)\\
 & 2013-02-27, 19:39:55 & -13.06, 75.07 & 00:40:40 &Setting (177)\\
 & 2013-02-27, 20:16:38 & 9.29, 249.51 & 12:54:54 &Rising (178)\\
 & 2013-02-27, 21:15:32 & -13.22, 51.07 &  00:39:39 &Setting (179)\\
 & 2013-02-27, 21:52:15 & 9.50, 225.46 & 12:54:54 &Rising (180)\\
\hline
\multirow{11}{*}{
\begin{tabular}{l}
Suzaku/XIS\\(408008010)
\end{tabular}}
 & 2013-09-16, 17:04:57 & -9.29, 106.58 & 00:11:11 &Rising (181)\\
 & 2013-09-16, 18:07:29 & 39.72, 289.45 & 13:25:25 &Setting (182)\\
 & 2013-09-16, 18:40:32 & -9.52, 82.69 & 00:11:11 &Rising (183)\\
 & 2013-09-16, 19:43:01 & 39.72, 265.48 & 13:24:24 &Setting (184)\\
 & 2013-09-16, 20:16:07 & -9.76, 58.81 & 00:11:11 &Rising (185)\\
 & 2013-09-16, 21:18:33 & 39.71, 241.51 & 13:24:24 &Setting (186)\\
 & 2013-09-16, 21:51:42 & -9.99, 34.92 & 00:11:11 &Rising (187)\\
 & 2013-09-16, 22:54:05 & 39.70, 217.54 & 13:24:24 &Setting (188)\\
 & 2013-09-17, 00:29:38 & 39.69, 193.64 & 13:24:24 &Setting (189)\\
 & 2013-09-17, 02:05:10 & 39.68, 169.67 & 13:23:23 &Setting (190)\\
 & 2013-09-17, 03:40:42 & 39.67, 145.69 & 13:23:23 &Setting (191)\\
\hline
\multirow{10}{*}{
\begin{tabular}{l}
Suzaku/XIS\\(108011010)
\end{tabular}}
 & 2013-09-30, 10:43:08 & -19.36, 191.56 & 23:29:29 &Rising (192)\\
 & 2013-09-30, 11:41:53 & 18.76, -4.22 & 11:25:25 &Setting (193)\\
 & 2013-09-30, 12:18:40 & -19.26, 167.58 & 23:28:28 &Rising (194)\\
 & 2013-09-30, 13:17:25 & 18.58, 331.72 & 11:24:24 &Setting (195)\\
 & 2013-09-30, 13:54:12 & -19.16, 143.59 & 23:28:28 &Rising (196)\\
 & 2013-09-30, 14:52:57 & 18.40, 307.66 & 11:23:23 &Setting (197)\\
 & 2013-09-30, 15:29:44 & -19.06, 119.60 & 23:28:28 &Rising (198)\\
 & 2013-09-30, 16:28:29 & 18.22, 283.61 & 11:22:22 &Setting (199)\\
 & 2013-09-30, 17:05:17 & -18.94, 95.67 & 23:27:27 &Rising (200)\\
 & 2013-09-30, 18:04:01 & 18.04, 259.55 & 11:22:22 &Setting (201)\\
 & 2013-09-30, 18:40:49 & -18.84, 71.69 & 23:27:27 &Rising (202)\\
 & 2013-09-30, 19:39:33 & 17.86, 235.50 & 11:21:21 &Setting (203)\\
 & 2013-09-30, 21:15:06 & 17.65, 211.48 & 11:21:21 &Setting (204)\\
 & 2013-09-30, 22:50:38 & 17.47, 187.42 & 11:20:20 &Setting (205)\\
 & 2013-10-01, 00:26:10 & 17.29, 163.36 & 11:19:19 &Setting (206)\\
 & 2013-10-01, 02:01:42 & 17.11, 139.30 & 11:18:18 &Setting (207)\\
 & 2013-10-01, 03:37:14 & 16.93, 115.25 & 11:18:18 &Setting (208)\\
 & 2013-10-01, 04:14:02 & -18.18, 287.81 & 23:25:25 &Rising (209)\\
 & 2013-10-01, 05:12:47 & 16.72, 91.24 & 11:17:17 &Setting (210)\\
 & 2013-10-01, 05:49:34 & -18.06, 263.83 & 23:24:24 &Rising (211)\\
 & 2013-10-01, 06:48:19 & 16.53, 67.18 & 11:17:17 &Setting (212)\\
 & 2013-10-01, 07:25:06 & -17.95, 239.83 & 23:24:24 & Rising (213)\\
 & 2013-10-01, 08:23:51 & 16.35, 43.12 & 11:16:16 &Setting (214)\\
 & 2013-10-01, 09:00:38 & -17.84, 215.84 & 23:23:23 &Rising (215)\\
 & 2013-10-01, 09:59:23 & 16.17, 19.06 & 11:15:15 &Setting (216)\\
\hline
\multirow{3}{*}{
\begin{tabular}{l}
Hitomi/HXI\\(100044010) 
\end{tabular}}
 & 2016-03-25, 14:52:42 & 32.73, 144.27 & 00:29:29 &Setting (217)\\
 & 2016-03-25, 16:28:43 & 32.63, 120.08 & 00:29:29 &Setting (218)\\
 & 2016-03-25, 18:02:24 & 35.27, 128.67 & 02:37:37 &Rising (219)\\
\end{longtable}

\clearpage

%{\small
\begin{table}
 \caption{Combined N and O column number densities measured with Suzaku/XIS}
 \label{tab:NOcol_suzaku}
\begin{tabular}{lccccc}
\hline
Altitude (km)  & \multicolumn{5}{c}{Column density (10$^{22}$\,cm$^{-2}$)} \\
& 106012010 & 106012010 & 106014010 & 106014010 & Mean for 10 epochs (SD)\\
& Setting & Rising & Setting & Rising & Setting $+$ Rising\\
\hline
200--210 & 1.6$\pm$0.6$\times$10$^{-4}$ & 0.9$\pm$0.4$\times$10$^{-4}$& 
1.8$\pm$0.5$\times$10$^{-4}$ & 2.9$\pm$0.9$\times$10$^{-4}$& 1.4$\times$10$^{-4}$ (0.7$\times$10$^{-4}$) \\
190--200 & 1.6$\pm$0.6$\times$10$^{-4}$& 1.1$\pm$0.4$\times$10$^{-4}$& 
1.8$\pm$0.6$\times$10$^{-4}$ &2.4$\pm$0.6$\times$10$^{-4}$ & 1.7$\times$10$^{-4}$ (0.7$\times$10$^{-4}$)\\
180--190 & 2.8$\pm$0.6$\times$10$^{-4}$& 2.0$\pm$0.5$\times$10$^{-4}$& 
3.0$\pm$0.6$\times$10$^{-4}$ & 3.3$\pm$0.9$\times$10$^{-4}$& 2.1$\times$10$^{-4}$ (0.7$\times$10$^{-4}$)\\
170--180 & 3.8$\pm$0.7$\times$10$^{-4}$& 2.0$\pm$0.5$\times$10$^{-4}$& 
4.0$\pm$0.7$\times$10$^{-4}$ & 4.3$\pm$0.7$\times$10$^{-4}$& 3.2$\times$10$^{-4}$ (1.1$\times$10$^{-4}$) \\
160--170 & 4.4$\pm$0.7$\times$10$^{-4}$ & 3.3$\pm$0.5$\times$10$^{-4}$  & 
4.6$\pm$0.7$\times$10$^{-4}$ & 4.8$\pm$1.0$\times$10$^{-4}$& 4.2$\times$10$^{-4}$ (1.4$\times$10$^{-4}$)\\
150--160 & 6.8$\pm$0.1$\times$10$^{-4}$ & 5.2$\pm$0.6$\times$10$^{-4}$& 
6.6$\pm$0.8$\times$10$^{-4}$ & 7.3$\pm$0.9$\times$10$^{-4}$& 6.3$\times$10$^{-4}$ (1.4$\times$10$^{-4})$ \\
140--150 & 1.0$\pm$0.1$\times$10$^{-3}$ & 0.7$\pm$0.1$\times$10$^{-3}$& 
1.1$\pm$0.1$\times$10$^{-3}$ & 1.1$\pm$0.1$\times$10$^{-3}$& 0.93$\times$10$^{-3}$ (0.22$\times$10$^{-3}$) \\
138--148 & 1.1$\pm$0.1$\times$10$^{-3}$ &0.8$\pm$0.1$\times$10$^{-3}$  &
1.1$\pm$0.1$\times$10$^{-3}$ & 1.1$\pm$0.2$\times$10$^{-3}$& 1.1$\times$10$^{-3}$ (0.3$\times$10$^{-3}$)\\
136--146 & 1.2$\pm$0.1$\times$10$^{-3}$ & 0.9$\pm$0.1$\times$10$^{-3}$&
1.3$\pm$0.1$\times$10$^{-3}$ & 1.3$\pm$0.2$\times$10$^{-3}$& 1.2$\times$10$^{-3}$ (0.3$\times$10$^{-3}$)\\
134--144 & 1.2$\pm$0.1$\times$10$^{-3}$& 1.0$\pm$0.1$\times$10$^{-3}$ &
1.4$\pm$0.1$\times$10$^{-3}$ & 1.4$\pm$0.1$\times$10$^{-3}$& 1.4$\times$10$^{-3}$ (0.3$\times$10$^{-3}$)\\
132--142 & 1.3$\pm$0.1$\times$10$^{-3}$ & 1.1$\pm$0.1$\times$10$^{-3}$ &
1.6$\pm$0.1$\times$10$^{-3}$ & 1.7$\pm$0.1$\times$10$^{-3}$& 1.6$\times$10$^{-3}$ (0.4$\times$10$^{-3}$)\\
130--140 & 1.4$\pm$0.1$\times$10$^{-3}$ & 1.3$\pm$0.1$\times$10$^{-3}$ & 
1.8$\pm$0.1$\times$10$^{-3}$ & 1.9$\pm$0.1$\times$10$^{-3}$ & 1.7$\times$10$^{-3}$ (0.4$\times$10$^{-3}$) \\
128--138 & 1.7$\pm$0.1$\times$10$^{-3}$ & 1.4$\pm$0.1$\times$10$^{-3}$&
2.0$\pm$0.1$\times$10$^{-3}$ & 2.0$\pm$0.1$\times$10$^{-3}$ & 1.9$\times$10$^{-3}$ (0.4$\times$10$^{-3}$) \\
126--136 & 2.0$\pm$0.1$\times$10$^{-3}$ & 1.6$\pm$0.1$\times$10$^{-3}$&
2.3$\pm$0.1$\times$10$^{-3}$ & 2.1$\pm$0.1$\times$10$^{-3}$& 2.2$\times$10$^{-3}$ (0.2$\times$10$^{-3}$)\\
124--134 & 2.4$\pm$0.2$\times$10$^{-3}$ & 1.8$\pm$0.1$\times$10$^{-3}$&
2.8$\pm$0.2$\times$10$^{-3}$ &2.3$\pm$0.2$\times$10$^{-3}$ & 2.6$\times$10$^{-3}$ (0.6$\times$10$^{-3}$)\\
122--132 & 2.6$\pm$0.2$\times$10$^{-3}$ & 2.0$\pm$0.1$\times$10$^{-3}$&
3.1$\pm$0.2$\times$10$^{-3}$ & 2.6$\pm$0.2$\times$10$^{-3}$& 3.0$\times$10$^{-3}$ (0.7$\times$10$^{-3}$) \\
120--130 & 2.9$\pm$0.2$\times$10$^{-3}$ & 2.2$\pm$0.1$\times$10$^{-3}$ & 
3.5$\pm$0.2$\times$10$^{-3}$ & 3.3$\pm$0.3$\times$10$^{-3}$  & 3.2$\times$10$^{-3}$ (0.8$\times$10$^{-3}$) \\
118--128 & 3.3$\pm$0.2$\times$10$^{-3}$ & 2.6$\pm$0.1$\times$10$^{-3}$ &
4.1$\pm$0.2$\times$10$^{-3}$ & 4.0$\pm$0.4$\times$10$^{-3}$& 3.9$\times$10$^{-3}$ (1.1$\times$10$^{-3}$) \\
116--126 & 3.8$\pm$0.3$\times$10$^{-3}$ & 3.1$\pm$0.1$\times$10$^{-3}$&
4.6$\pm$0.3$\times$10$^{-3}$ & 4.4$\pm$0.4$\times$10$^{-3}$& 4.5$\times$10$^{-3}$ (1.2$\times$10$^{-3}$)\\
114--124 & 4.8$\pm$0.3$\times$10$^{-3}$ & 3.6$\pm$0.2$\times$10$^{-3}$&
5.8$\pm$0.4$\times$10$^{-3}$ & 6.5$\pm$0.5$\times$10$^{-3}$& 5.3$\times$10$^{-3}$ (1.5$\times$10$^{-3}$) \\
112--122 & 6.2$\pm$0.4$\times$10$^{-3}$ & 4.6$\pm$0.2$\times$10$^{-3}$&
6.1$\pm$0.3$\times$10$^{-3}$ & 7.2$\pm$0.5$\times$10$^{-3}$& 6.3$\times$10$^{-3}$ (1.7$\times$10$^{-3}$)\\
110--120 & 7.7$\pm$0.4$\times$10$^{-3}$ & 5.2$\pm$0.3$\times$10$^{-3}$& 
8.0$\pm$0.5$\times$10$^{-3}$ & 8.4$\pm$0.6$\times$10$^{-3}$ & 7.8$\times$10$^{-3}$ (2.0$\times$10$^{-3}$) \\
108--118 & 8.7$\pm$0.5$\times$10$^{-3}$ & 6.9$\pm$0.3$\times$10$^{-3}$&
10.3$\pm$0.6$\times$10$^{-3}$ & 11.1$\pm$0.7$\times$10$^{-3}$& 11.3$\times$10$^{-3}$ (3.3$\times$10$^{-3}$)\\
106--116 & 9.7$\pm$0.5$\times$10$^{-3}$ & 9.2$\pm$0.4$\times$10$^{-3}$ &
13.4$\pm$0.9$\times$10$^{-3}$ & 13.1$\pm$0.9$\times$10$^{-3}$& 13.9$\times$10$^{-3}$ (4.5$\times$10$^{-3}$)\\
104--114 & 0.015$\pm$0.001 & 0.011$\pm$0.001&
0.016$\pm$0.001 & 0.018$\pm$0.001 & 0.018 (0.005) \\
102--112 & 0.024$\pm$0.002 & 0.016$\pm$0.001&
0.023$\pm$0.002 & 0.018$\pm$0.002 & 0.023 (0.006) \\
100--110 & 0.044$\pm$0.004 & 0.019$\pm$0.001 & 
0.032$\pm$0.002 & 0.024$\pm$0.003 & 0.030 (0.008) \\
90--100 & 0.20$\pm$0.03 & 0.12$\pm$0.01 &  
0.24$\pm$0.04 & 0.20$\pm$0.03 & 0.18 (0.07) \\
\hline
\end{tabular}
% \multicolumn{5}{l}{\small\it $^a$Mean and standard deviation for the 10 Suzaku data sets.}\\
% \endfoot
%\tablenotetext{a}{}
\end{table}
%}

\begin{table}
 \caption{Combined N and O column number densities measured with Hitomi/HXI}
 \label{tab:NOcol_hitomi}
\begin{tabular}{lccc}
\hline
Altitude (km) & \multicolumn{3}{c}{Column density (10$^{22}$\,cm$^{-2}$)} \\
& Setting & Rising & Mean (SD)\\
\hline
100--110 & 0.018$\pm$0.005 & 0.019$\pm$0.008 & 0.018 (0.001) \\
90--100 & 0.12$\pm$0.07 & 0.20$\pm$0.02 & 0.16 (0.07) \\
80--90 & 0.64$\pm$0.03 & 1.15$\pm$0.08 & 0.89 (0.25)\\
70--80 & 4.1$\pm$0.3 & 7.0$\pm$0.9 & 5.5 (2.0) \\
\hline
\end{tabular}
% \multicolumn{7}{l}{\small\it $^a$Mean and standard deviation for the 10 Suzaku data sets.}\\
% \endfoot
%\tablenotetext{a}{}
\end{table}

\begin{table}
 \caption{Trend of the N$+$O column density}
 \label{tab:NOcol_trend}
\begin{tabular}{lcc}
\hline
Altitude & \multicolumn{2}{c}{Rate} \\
(km) & (10$^{16}$\,cm$^{-2}$\,yr$^{-1}$) & (\%\,yr$^{-1}$)\\
\hline
190--200 & 8$\pm$9 & 9.4$\pm$11.3\\
180--190 & 10$\pm$9 & 9.4$\pm$8.5 \\
170--180 & 15$\pm$9 & 7.8$\pm$4.7 \\
160--170 & 34$\pm$12 & 16.1$\pm$5.6 \\
150--160 & 13$\pm$13 & 2.7$\pm$2.7 \\
140--150 & 26$\pm$17 & 3.7$\pm$2.3 \\
130--140 & $-1\pm$22 & $-0.1\pm$1.6 \\
120--130 & 5$\pm$34 & 0.19$\pm$1.3 \\
110--120 & $-110\pm$85 & $-1.7\pm$1.4 \\
100--110 & 730$\pm$400 & 4.2$\pm$2.3 \\
90--100 & 1600$\pm$2300 & 1.4$\pm$2.1 \\
\hline
\end{tabular}
% \multicolumn{7}{l}{\small\it $^a$Mean and standard deviation for the 10 Suzaku data sets.}\\
% \endfoot
%\tablenotetext{a}{}
\end{table}

%% SIDEWAYS FIGURE and TABLE
% AGU prefers the use of {sidewaystable} over {landscapetable} as it causes fewer problems.
%

\end{document}